\documentclass[preprint,superscriptaddress]{revtex4}

\usepackage{setspace}
\usepackage{times}

\usepackage{amsmath}

\usepackage{graphicx}

\usepackage{layout}

\usepackage{mathrsfs}

\usepackage{caption} 
\begin{document}
\verb| |\\   

{\large
\begin{center}
\protect{\vspace{-2cm}}
Jefferson Lab PAC 46 Letter of Intent
\end{center}
}

\title{Accessing DEMP and DVCS at Backward Angles above \\the Resonance Region}

\author{W. B. Li}
\affiliation{College of William and Mary, Williamsburg, VA}
\author{J. R. Stevens}
\affiliation{College of William and Mary, Williamsburg, VA}
\author{G. M. Huber}
\affiliation{University of Regina, Regina, SK  Canada}

\date{\today}

\begin{abstract}

The proposed measurement is a dedicated study to investigate the exclusive electroproduction process: $^1$H$(e, e^{\prime}p)X$, in the backward angle above the resonance region. Here, the produced particle $X$ ($\pi^0$ or $\gamma$) is emitted 180 degrees opposite to the virtual photon momentum. This study will apply the well known L/T separation method of the electroproduction process to this unexplored backward angle kinematics region. 

The available theoretical frameworks give parallel interpretations to the backward angle meson production at the proposed kinematics. According to the QCD GPD-like TDA model, backward meson production as the virtual-photon probes the transverse meson cloud structure inside of the nucleon; whereas the hadronic Regge based model describes backward meson production as the interference between nucleon exchange and the meson produced via re-scattering of the nucleon. Testing these two approaches is like testing the onset of quarks and gluon physics, i.e. going deeper in the structure of nucleon to probe its quark and gluon content. Combining knowledge from both frameworks has the potential to provide complementary knowledge to the forward angle physics programs and obtain new physics insights to study the QCD transition from meson-nucleon to quark-hadron degrees of freedom.

\end{abstract}

\maketitle
\clearpage

\section{Introduction}

\begin{figure}
\centering
\includegraphics[width=0.8\textwidth]{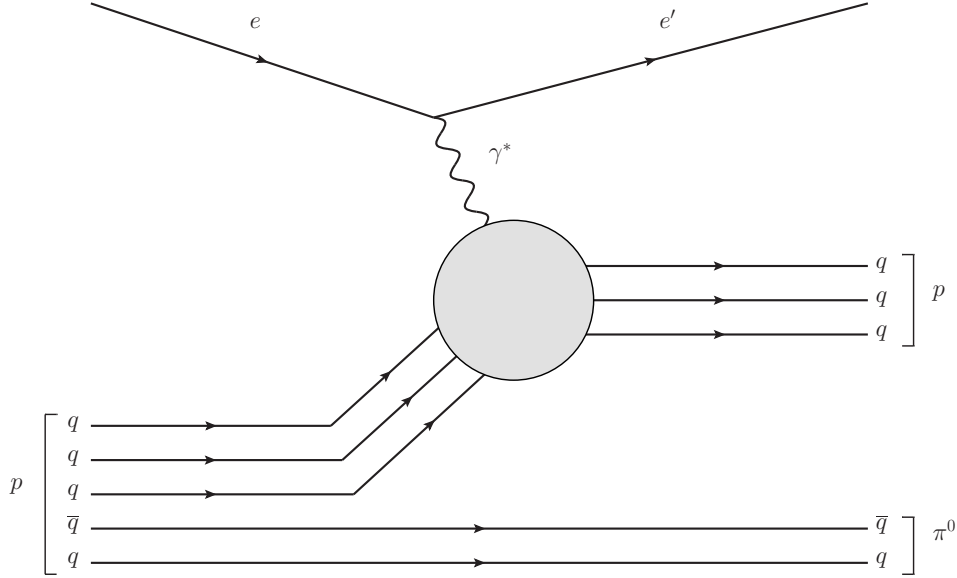}
\caption{Cartoon demonstration of a ``knocking a proton out of a proton process'' above the resonance region ($\sqrt{s}=W>2$~GeV)~\cite{weiss17}. In this case, a backward $\pi^0$ is produced.}
\label{fig:pi0_cartoon}
\end{figure}

In this letter of intent, we present a unique opportunity to access Deep Exclusive Meson Production (DEMP) and Deeply Virtual Compton Scattering (DVCS) in the backward angle regime.  These reactions have the advantage of being detected simultaneously. In the case of DEMP, the primary experimental observable involves the exclusive $\pi^0$ electroproduction: $^1$H$(e, e^{\prime}p)\pi^{0}$, with a kinematic coverage of $ 2 < Q^2 < 6$~GeV$^2$ at fixed $x_{\rm B}=0.36$ and $W>2$~GeV. Since the $\pi^0$ is produced almost at 180$^\circ$ opposite to the direction of the virtual photon momentum (corresponding to extreme backward angles), the Mandelstam variable for crossed four-momentum transfer squared $u^{\prime} = u-u_{\textrm{min}} = 0$~GeV$^2$ ($u$-channel skewness $\xi_u\sim1$). At selected $Q^2$ settings, the full L/T/LT/TT cross section separation will be performed. The backward DVCS events, $^1$H$(e, e^{\prime}p)\gamma$, will be fortuitously detected as part of the physics background. Due to its unusual kinematics, the backward angle reaction is often referred to as a ``knocking a proton out of a proton process'', as shown in Fig.~\ref{fig:pi0_cartoon}.


The proposed $\pi^0$ production measurement (demonstrated in Fig.~\ref{fig:pi0_cartoon}) uses the standard Hall~C equipment, standard-gradient unpolarized electron beam and liquid hydrogen (LH$_2$) target. Since the missing mass reconstruction method does not require a detection of the produced meson, this permits access to a unique backward angle kinematics region that was previously unexplored. The L/T separation technique is the same as was used successfully by many previous Hall A and C experiments during the 6~GeV era of CEBAF, a successful example being the pion form factor experiment~\cite{volmer01, blok08}.

Compared to the $\pi^0$ events cleanly identifying  the DVCS events is more complicated, requiring an additional detector such as the Neutral Particle Spectrometer (NPS)~\cite{tanja15}, currently under construction.  This detector must be used to detect the backward-scattered $\gamma$, which is further discussed in Sec.~\ref{sec:DVCS}. 

The proposed measurements in this Letter of Intent have a threefold motivation:
\begin{itemize}

\item \textbf{Establishing a systematic program to probe parton-like structure within the nucleon through a QCD GPD-like model: the baryon-to-meson Transition Distribution Amplitude (TDA)~\cite{lansberg07}, and challenging the two specific predictions made by TDA} (see Sec.~\ref{sec:tda_prediction}).

\item \textbf{Extending the $-t$ coverage of forward angle physics program.} Introduction of the $Q^2$ dependence to the hadronic Regge based model by J. M. Laget~\cite{laget00, laget02, laget04}, inspired studies of experimental cross section scaling (falling) with respect to $-t$ and became one of the predominant methodologies for accessing  a wide range of hadronic observables within a consistent theoretical framework. However, electroproduction in the extreme backward angle region (corresponding to a region of ultra high $-t$ or small $-u$) has not been explored. The backward angle observables, in combination with the forward angle observables, give the full $-t$ evolution picture, which is important for Regge phenomenological studies.
 

\item \textbf{Comparing the effectiveness of the hadronic Regge based and TDA models, qualitatively studying the transition of QCD from meson-nucleon to quark-gluon degrees of freedom.}  It is anticipated that as the $Q^2$ is extended towards the optimal range of the TDA framework of $Q^2$ $>$ 10~GeV$^2$, the Regge-based model might become less effective due to the transition from the hadronic to partonic degrees of freedom of the nucleon. Studying the ``crossing point'' in terms of model effectiveness between the hadronic Regge based (exchanges of mesons and baryons) and TDA (exchanges of quarks and gluons), is equivalent to studying the QCD transition.  

\end{itemize}

\section{Summary on Backward Angle Physics from JLab 6~GeV}

\label{sec:exp_summary}

At Jefferson Lab, direct or indirect measurements of exclusive meson electroproduction at large scattering angles are not a new concept.  Here, indirect measurement implies the usage of the missing mass reconstruction technique. During the 6~GeV era, there have been a few examples of such studies. In this section, we present a short overview to some of the important pioneering studies of backward angle physics.


\subsection{Backward VCS and $\pi^0$ Electroproduction at Hall A}

\begin{figure}[t]
\centering
\includegraphics[width=0.90\textwidth]{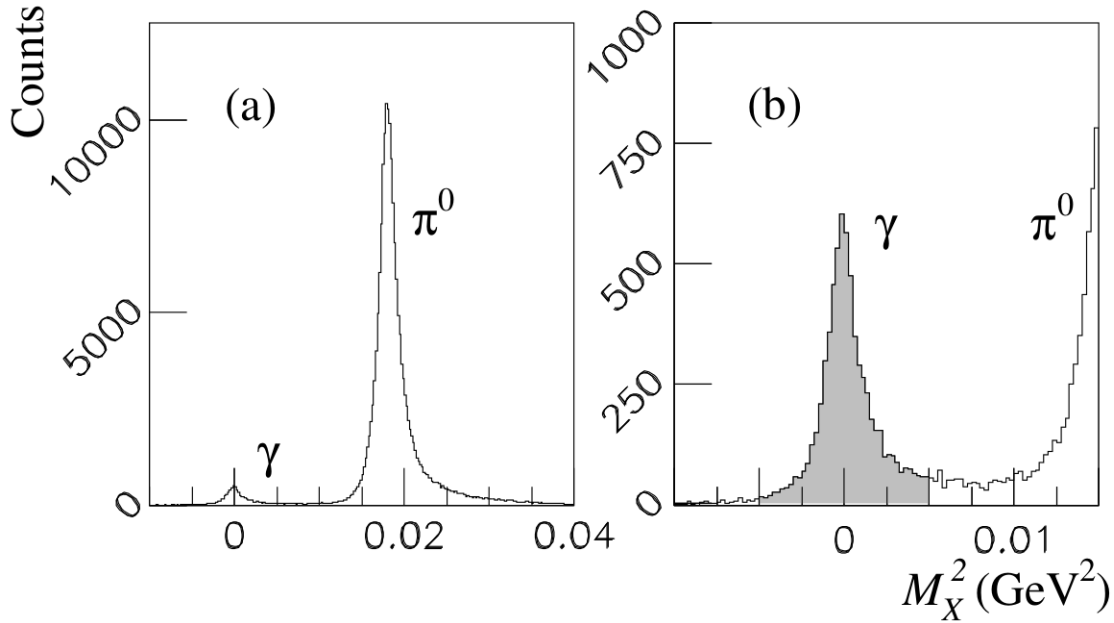}
\caption{Squared missing mass $M^2_X$ for an experimental setting $W=1.2$~GeV is shown in plot (a). The zoomed distribution around $\gamma$ peak is shown in (b). These plots were published in Ref.~\cite{laveissiere09}.}
\label{fig:mx2}
\end{figure}

Since the early stage of JLab (1993), backward angle $^1$H$(e, e^{\prime}p)\gamma$ and $^1$H$(e, e^{\prime}p)\pi^0$ were attempted by a dedicated Hall A experiment E93-050~\cite{audit93}, and later by E00-110~\cite{chen00} in the nucleon resonance region. Both experiments used the 4~GeV electron beam colliding with a liquid hydrogen target, where a pair of High Resolution Spectrometers (HRSs) were used to detect the scattered electron and proton in coincidence.  The forward-going proton was detected in parallel kinematics and the `recoil' $\pi^0,~\gamma$ emitted at backward angle at low momentum.  The missing mass reconstruction technique was used to reconstruct the final state $\gamma$ as well as $\pi^0$ events.  An example of the reconstructed missing mass squared distribution from E00-110 is shown in Fig.~\ref{fig:mx2}.

E93-050 and E00-110 had a common physics objective, which was to access the Compton photon scattered at backward angles in the nucleon resonance region ($S_{11}$ and $D_{13}$), whereas the $\pi^0$ was detected as the dominant background. Thanks to the good particle momentum resolution of the HRSs, separating the $\gamma$ and $\pi^0$ peaks was a relatively easy task (as shown in Fig,~\ref{fig:mx2}).  However, a clean separation of the Virtual Compton Scattering (VCS) events from Bethe-Heitler contribution in the resonance region was a challenging task, and E93-050 only published the $\pi^{0}$ production cross section~\cite{laveissiere04}.


In 2008, Laveissiere, et al., published the first measurement of the backward angle VCS cross section with the data from E00-110~\cite{laveissiere09}. This experiment was performed in the nucleon resonance region and had $W = 1.5$ GeV, $Q^2 =1$ GeV$^2$. These early measurements demonstrated the feasibility of extracting the $^1$H$(e, e^{\prime}p)\gamma$ and $^1$H$(e, e^{\prime}p)\pi^0$ cross sections in the backward region.

Despite their different physics motivation, the experimental technique used by E93-050 and E00-100 are quite similar to the one proposed in this LOI. The relative height and width of $\gamma$ and $\pi^0$ peaks from these previous measurements are useful benchmarks for the cross section estimation and resolution requirements.

\subsection{High $-t$ charged $\pi$ Electroproduction at Hall B}
\label{sec:clas}


The CLAS detector, in comparison to the Halls A and C spectrometers, presents the great advantage of a wide angular acceptance. Since the cross section for a given electroproduction reaction falls exponentially as a function of $-t$ (a larger $-t$ value corresponds to a wider scattering angle), it is difficult to determine the detector efficiency for wide scattering angles. After years of careful study, K. Park et al. published results for exclusive $\pi^+$ electroproduction, $^1$H$(e, e^\prime \pi^+)n$, near the backward angle above the resonance region~\cite{park18}. The $Q^2$ coverage is $1.5< Q^2 < 4.5$~GeV$^2$, at $W\sim$2.2 GeV, $-u =$0.5 GeV$^2$. 

\begin{figure}[t]
\centering
\includegraphics[width=0.80\textwidth]{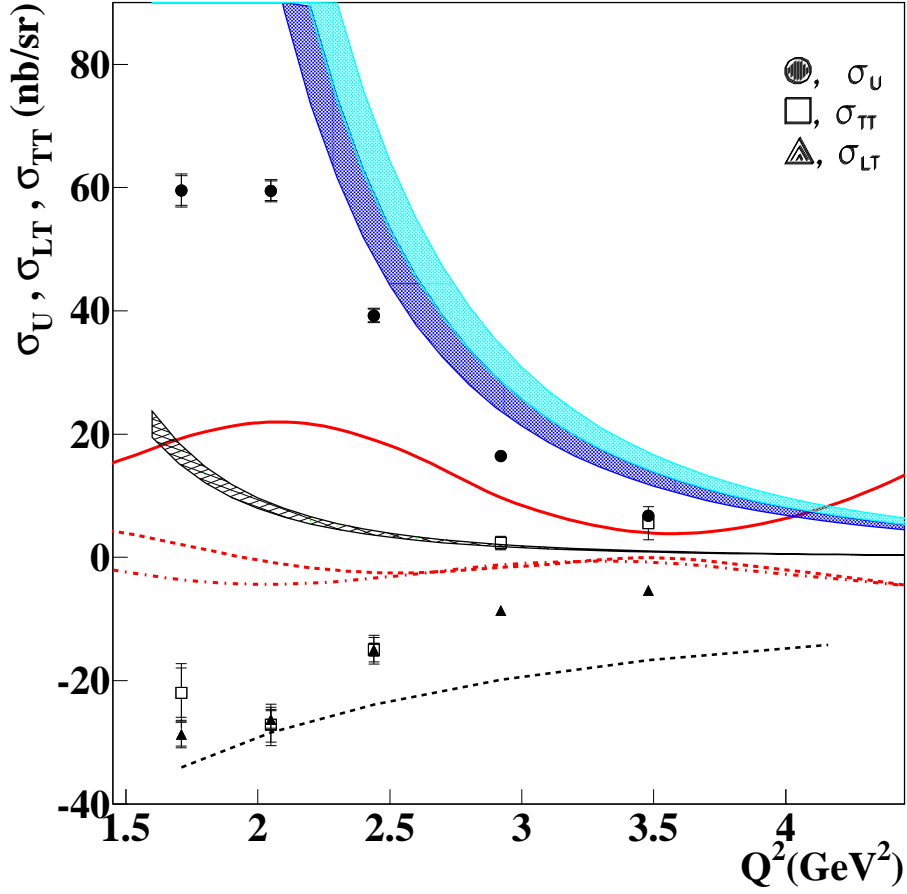}
\caption{The structure functions $\sigma_{u}$, $\sigma_{\rm TT}$ and $\sigma_{\rm LT}$ as a function of $Q^2$. The bands refer to model calculations of $\sigma_{u}$ in the TDA description with different nucleon DA models; dark blue band: COZ~\cite{chernyak89} $N$ DA model, light blue band: KS~\cite{king87}, black band: BLW NNLO~\cite{lenz09}. This plot was published in Ref.~\cite{park18}.} 
\label{fig:park18}
\end{figure}



The publication of this result was an important step for $u$-channel physics. This result presented indications of $Q^2$-scaling (particularly for $Q^2>2$ GeV$^2$), consistent with the prediction of the QCD GPD-like TDA factorization scheme at a much lower $Q^2$ range than originally expected.  This is demonstrated by the close agreement between the blue TDA band and the unseparated $\sigma_U$ in Fig.~\ref{fig:park18}.


\subsection{Backward $\omega$ Electroproduction at Hall C}

\label{sec:omega}

The recently completed (2017) Ph.~D thesis work (from Hall C~\cite{wenliang17}) demonstrated that the missing mass reconstruction technique, in combination with the high precision spectrometers in coincidence mode at Hall C, can be used to reliably extract the backward-angle $\omega$ cross section through the exclusive reaction $^1$H$(e, e^{\prime}p)\omega$, while performing a full L/T separation. The experiment has central $Q^2$ values of 1.60 and 2.45 GeV$^2$, at $W = 2.21$ GeV. There was significant coverage in $\phi$ and $\epsilon$, which allowed separation of $\sigma_{\textrm{T,L,LT,TT}}$. The data set has a unique $u$ coverage near $-u \sim 0$, which corresponds to $-t > 4$ GeV$^2$.

\begin{figure}[htp!]
\centering
\includegraphics[width=0.65\textwidth]{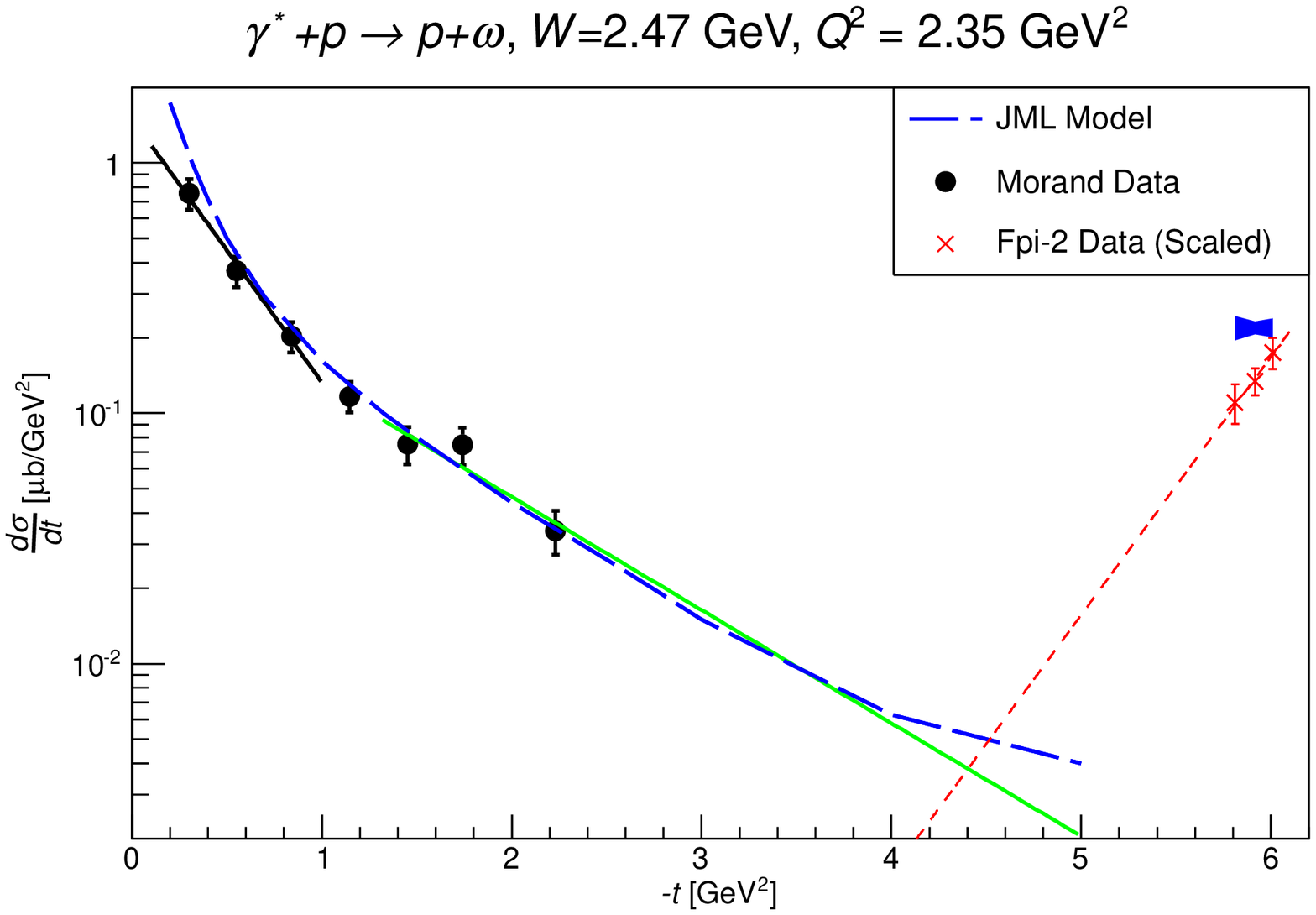}
\caption{Unseparated differential cross section, $d\sigma_{u}/dt$ versus $-t$ for $W$ = 2.47~GeV, $Q^2$ = 2.35~GeV. The black dots are published CLAS results~\cite{morand05}. The red crosses are reconstructed $\sigma_u$ using $\sigma_{\rm T}$ and $\sigma_{\rm L}$ from Hall C (scaled to same kinematics) \cite{wenliang17}, the systematic error bands are shown in blue. The blue dashed line represents the prediction of the hadronic Regge based model~\cite{laget04}. The black line is a fitted curve showing the contribution of the forward angle soft process (e.g. meson exchange); the green solid line is a fitted curve showing a flatter $-t$ dependence due to a harder process interaction; red dashed line is a fitted curve which might indicate the contribution due to the softer baryon exchange in the backward angle. This plot was published in Ref.~\cite{wenliang17}.} 
\label{fig:omega}
~\\[5mm]
\includegraphics[width=0.625\textwidth]{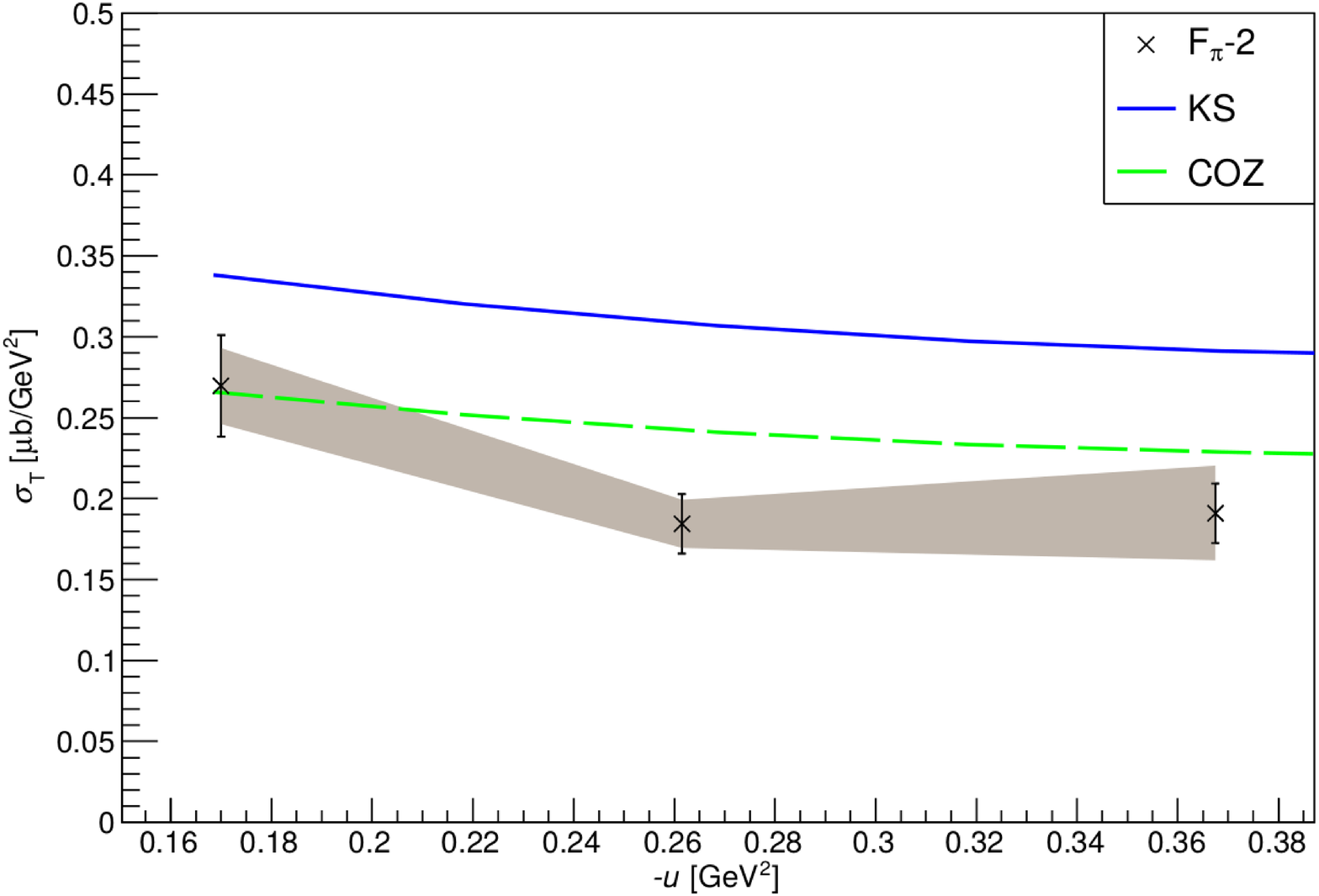}
\caption{ $\sigma_{\textrm{T}}$ versus $-u$ for $Q^2=2.45$~GeV$^2$. The blue solid and green dashed lines represent the TDA calculation \cite{pire15} using the KS~\cite{king87} and COZ~\cite{chernyak89} nucleon DA models, respectively. This plot was published in Ref.~\cite{wenliang17}.} 
\label{fig:sigT}
\end{figure}

The extracted cross sections (red crosses) show evidence of a backward angle peak for $\omega$ exclusive electroproduction; an example setting of $Q^2=2.35$ GeV$^2$ is shown in Fig~\ref{fig:omega}. Note that the forward angle ($t$-channel) peak from the CLAS-6 data~\cite{morand05} is also shown. Previously, this phenomenon showing both forward and backward angle peaks was only observed in the meson photoproduction data~\cite{vgl96, guidal97}.  

\begin{figure}[t]
\centering
\includegraphics[width=0.75\textwidth]{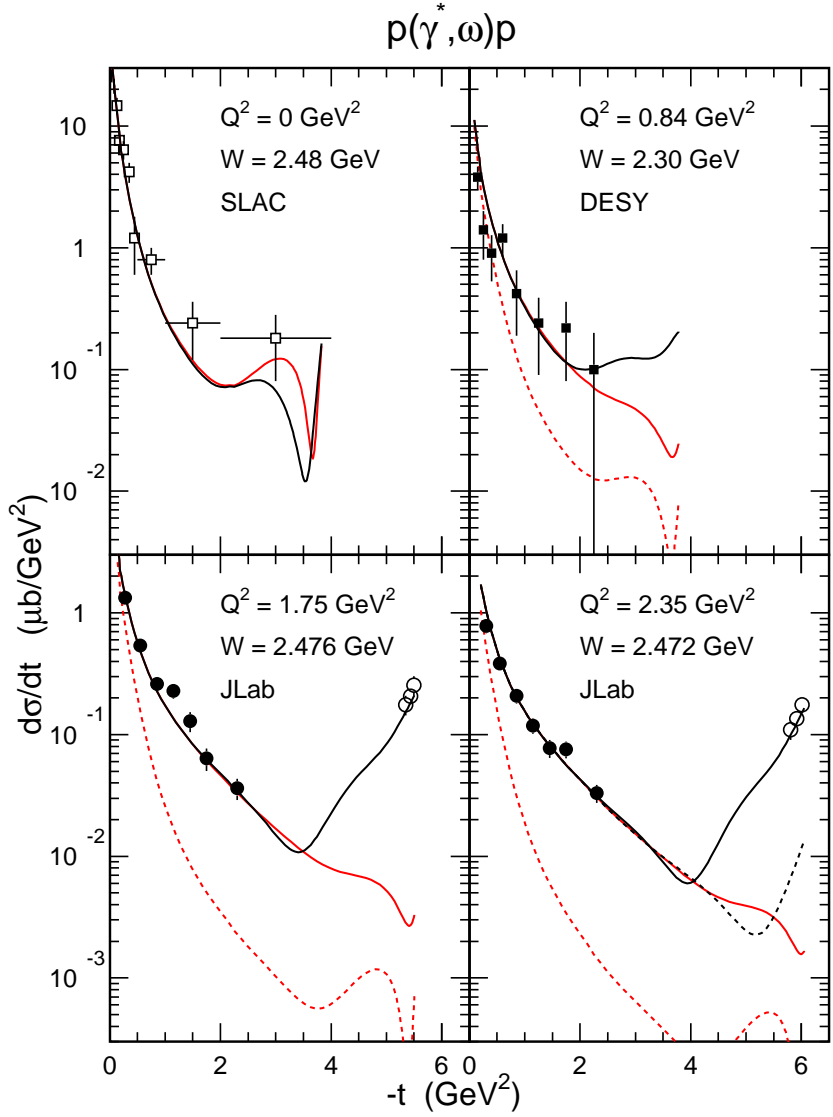}
\caption{The cross section evolution for exclusive $\omega$ meson production as function a function of $-t$. Note that the bottom panels show $\omega$ electroproduction data from CLAS (dots) and Hall C (circles), the $Q^2 = 1.75$ GeV$^2$ on the left and $Q^2 = 2.35$ GeV$^2$ on the right. The dashed red curves are the predictions of this basic model when a constant cutoff mass is used in the meson electric and magnetic form factors~\cite{laget04}. The full red curves are the predictions when a $t$-dependent cutoff mass is used~\cite{laget04}. The black dashed line is the prediction of the nucleon degenerated pole only. The full line curves take into account the interference between nucleon exchange and the $\omega$ produced via nucleon exchange re-scattering on the nucleon. Plot provided by J. M. Laget through private communication~\cite{laget18}.} 
\label{laget} 
\end{figure}

The $p(e,e^{\prime}p)\omega$ $d\sigma_{\rm T}/dt$ from Hall C \cite{wenliang17} are shown for an example setting at $Q^2=2.45$ GeV versus $-u$ are compared to the TDA model prediction \cite{pire15} in Fig.~\ref{fig:sigT}.  The TDA model predictions are within 1-2~$\sigma$ band of the data, depending on whether the COZ or KS nucleon distribution amplitudes (DA) are used. In addition, the indication of $\sigma_{\rm T}$ dominance over $\sigma_{\rm L}$ at $Q^2$ = 2.45 GeV$^2$, seems to agree with the postulated TDA factorization condition \cite{pire05}. As the JLab 12~GeV experiments can reach higher $Q^2$ values, the TDA formalism must be carefully studied and tested.

In a recent private communication with J. M. Laget~\cite{laget18}, he has shown that the hadronic Regge based model also gives a satisfactory description of the backward angle $\omega$ peak at both $Q^2$ settings, as shown in Fig.~\ref{laget}. In addition to the degenerated nucleon exchange amplitude, the hadronic Regge based model requires an elastic re-scattering cutoff, where the omega produced via nucleon exchange re-scatters on the nucleon. This interference leads to the black full line curves. The black dashed line is the prediction of the nucleon degenerated pole only~\cite{laget18}.

\subsection{Summary and Advantages for Backward $\pi^0$ Production}

In comparison to the backward $\omega$ or $\eta$ electroproduction processes, the reconstructed missing mass distribution for $\pi^0$ has little physics background underneath its narrow peak. This significantly reduces the complication associated with the background removal during the analysis. In addition, $\pi^0$ production has been a popular candidate for the theory studies~\cite{laget04, lansberg07}.  All these features make it a prime choice to initiate the backward-angle studies in the JLab 12~GeV era.  In addition, backward $\pi^0$ production has received significant interest beyond the JLab physics program and will be studied by the $\overline{\rm P}$ANDA experiment at FAIR~\cite{panda15} through the complementary process $\overline{p}+p\rightarrow\gamma^*+\pi^0$.






\section{Theoretical Context for Backward Angle $\pi^0$ electroproduction}


In the form $e$-$p$ scattering representation, the exclusive $\pi^0$ electroproduction $^1$H$(e, e^{\prime}p)\pi^{0}$ can be written as
\begin{equation}
e(k) + p(p_1) \rightarrow e^{\prime}(k^\prime) + \pi(p_{\pi}) + p^\prime(p_2)\,.
\end{equation}
If the virtual photon is considered as the projectile, then
\begin{equation}
\gamma^*(q) + p(p_1) \rightarrow \pi(p_{\pi}) + p^\prime(p_2)\,.
\label{eqn:reaction}
\end{equation}
Here, $p$ and $p^\prime$ are the proton target before and after the interaction; $e$ and $e^\prime$ are the electron before and after the interaction; $\gamma^{*}$ is the space-like virtual photon. The associated four-momentum for each particle is given inside of the bracket. The standard definition of the Mandelstam variables are defined as
\begin{equation}
s=(p_1 + q)^2; ~~~~ u =(p_\pi - p_1)^2; ~~~~~ t= (p_2 -p_1)^2.
\end{equation}
In the case of the forward-angle ($t$-channel) meson production process, the $\pi^0$ is produced in the same direction as the virtual photon momentum $q$ (known as the $q$-vector), and $-t\rightarrow 0$ (i.e. parallel kinematics). Correspondingly, the backward angle ($u$-channel) process produces $\pi^0$ in the opposite direction as the $q$-vector, and $-u \rightarrow -u_{min}$ (anti-parallel kinematics).


In the different kinematic regions, the backward meson production can be explained using different nucleon structure models. When the process is within the resonance region ($W < 2$~GeV), the $u$-channel process can be described using the nucleon fragmentation model which has a mild $Q$ dependence~\cite{weiss17}; when above the resonance region ($W > 2$~GeV), a more complicated parton based model is required to describe the $Q^n$ dependence. The latter is the research interest of this LOI.

Within the 6 GeV JLab kinematics coverage: $ W > 2$ GeV, $Q^2 < 3$ GeV, $x_{\rm B}=0.36$, there are two independent models that are capable of describing the existing data in the backward angle. The first is a hadronic Regge based model known as the JML model~\cite{laget04,laget18}, that explores meson-nucleon dynamics of hadron production reactions; the other model is a QCD GPD-like model known as the TDA~\cite{lansberg07} which offers direct description of the individual partons within the nucleon. In this section, we introduce how a backward-angle $\pi^0$ is produced according to both models and describe the benefits for studying them. Note that both models have the capability of calculating L/T separated cross sections and the leading twist TDAs predict $\sigma_{\textrm L}\sim0$~\cite{lansberg07}.





\subsection{The QCD Approach}
\label{sec:tda}


Generalized parton distributions (GPDs) are an improved description of the complex internal structure of the nucleon, which provide access to the correlations between the transverse position and longitudinal momentum distribution of the partons in the nucleon. In addition, GPDs give access to the orbital momentum contribution of partons to the spin of the nucleon~\cite{ji97, jo12}.

\begin{figure}[t]
\centering
\includegraphics[width=0.49\textwidth]{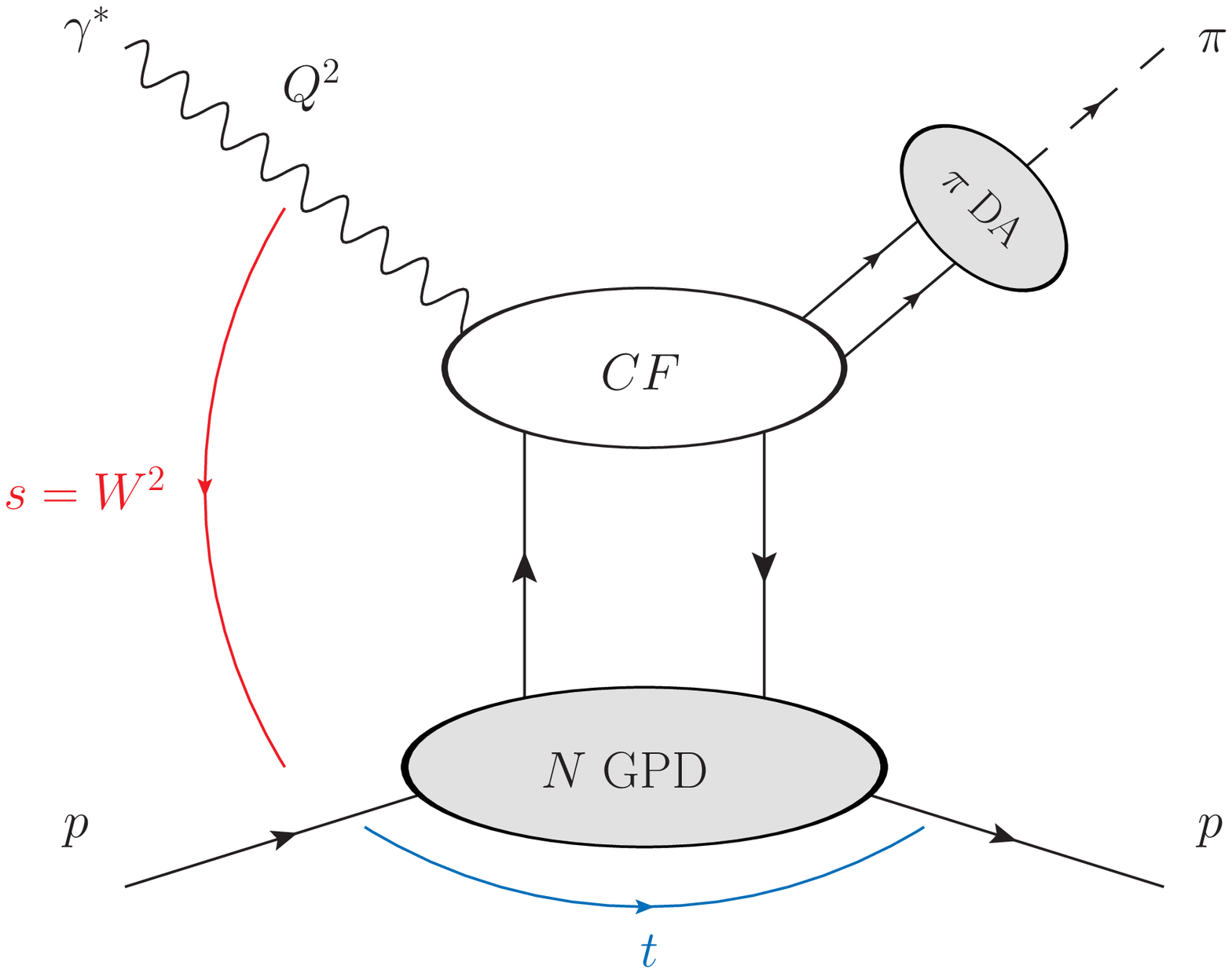}
\includegraphics[width=0.49\textwidth]{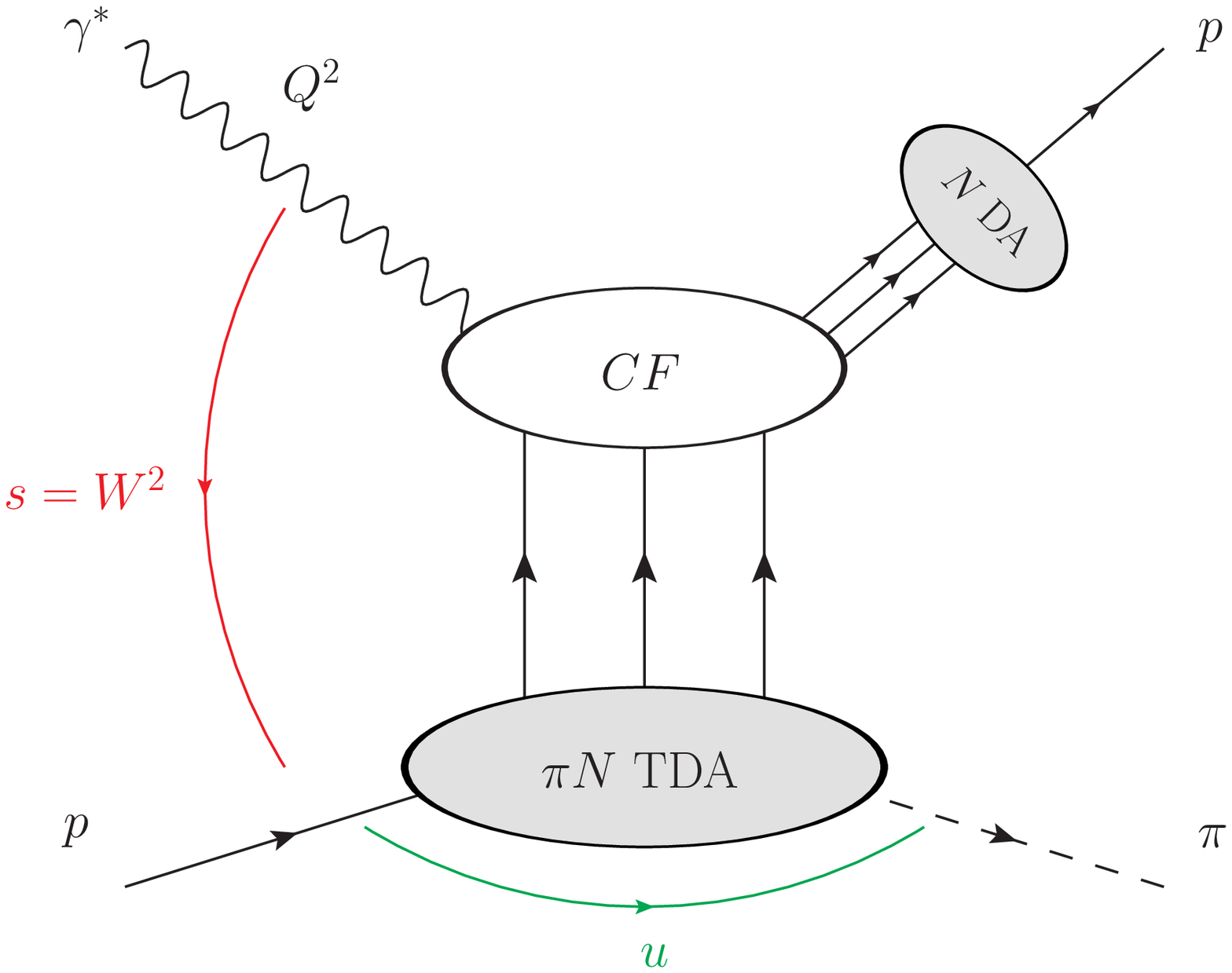}

\caption{(a) shows the $\pi^0$ electroproduction production interaction ($\gamma^*p\rightarrow p\pi^0$) diagram under the (forward-angle) GPD colinear factorization regime (large $Q^2$, large $s$, fixed $x_{\rm B}$, fixed $t\sim0$). $N$ GPD is the quark nucleon GPD (note that there are also gluon GPD that is not shown). $\pi$ DA stands for the vector meson distribution amplitude. The CF corresponds to the calculable hard process amplitude.  (b) shows the (backward-angle) TDA colinear factorization regime (large $Q^2$, large $s$, fixed $x_{\rm B}$, $u\sim0$) for $\gamma^*p\rightarrow  p \pi^0$. The $\pi N$ TDA is the transition distribution amplitude from a nucleon to a vector meson. These plots were created based on the original ones published in Ref.~\cite{lansberg11}. }

\label{fig:GPD_TDA}
\end{figure}

Currently, there is no known direct experimental access to the information encoded in GPDs~\cite{ji04}. The prime experimental channels for studying the GPDs are through the DVCS and DEMP processes~\cite{ji97}. Both processes rely on the colinear factorization (CF) scheme~\cite{collins97,radyushkin87}.  An example DEMP reaction, $\gamma^*p\rightarrow p\pi^0$, is shown in Fig.~\ref{fig:GPD_TDA}(a). In order to access the forward angle GPD colinear factorization regime ($\gamma^*p\rightarrow p\pi^0$ interaction), the kinematics variables requirements are as follows: sufficiently high $Q^2$, large $s$, fixed $x_{\rm B}$ and $t\sim0$~\cite{ji04, pire15}. Here, the definition of ``sufficiently high $Q^2$'' is process dependent terminology. Based on the existing DIS data~\cite{camacho06, girod08}, the GPD physics has shown that the range of ``sufficiently high $Q^2$'' ranges 1 to 5 GeV$^2$, this is sometimes referred to as the early scaling~\cite{pire18, voutier09}.

Under the colinear factorization regime, a parton is emitted from the nucleon GPDs ($N$ GPDs) and interacts with the incoming virtual photon, then returns to the $N$ GPDs after the interaction~\cite{ji04}. Studies~\cite{kroll16, liuti10} have shown that perturbative calculation methods can be used to calculate the CF process (top oval in Fig.~\ref{fig:GPD_TDA} (a)) and extract GPDs through factorization, while preserving the universal description of the hadronic structure in terms of QCD principles.

%
%

TDAs are the backward analog of GPDs, with their full name being the baryon-to-meson transition distribution amplitude ($\pi N$ TDA). TDAs describe the underlying physics mechanism of how the target proton transitions into a $\pi$ meson in the final state, shown in the gray oval in Fig.~\ref{fig:GPD_TDA}(b). One fundamental difference between GPDs and TDAs is that the TDAs require three parton exchanges between $\pi N$ TDA and CF.

Relevant to this discussion is the definition of skewness. For forward angle kinematics, in the regime where the handbag mechanism and GPD description may apply, the skewness is defined in the usual manner,
\begin{equation}
\xi_t=\frac{p_1^+-p_2^+}{p_1^++p_2^+},
\label{eqn:xi_t}
\end{equation}
where $p_1$, $p_2$ refer to the light-cone plus components of the initial and final proton momenta in Eqn.~\ref{eqn:reaction}, calculated in the CM frame \cite{kroll04}.  The subscript $t$ has been added to indicate that this skewness definition is typically used for forward-angle kinematics, where $-t\rightarrow -t_{min}$.  In this regime, $\xi_t$ is related to Bjorken-$x$ ($x_{\rm B}$), and is approximated by $\xi_t=x_{\rm B}/(2-x_{\rm B})$, up to corrections of order $1/Q^2$ \cite{favart15}.  This relation is an accurate estimate of $\xi_t$ to the few percent level for forward angle electroproduction.  However, for the backward angle kinematics of interest in this LOI, the approximate formula does not hold, and only Eqn.~\ref{eqn:xi_t} can be used.

In backward angle kinematics, where $-t\rightarrow -t_{max}$ and $-u\rightarrow -u_{min}$, the skewness is defined with respect to $u$-channel momentum transfer in the TDA (Transition Distribution Amplitude) formalism \cite{lansberg07},
\begin{equation}
\xi_u=\frac{p_1^+-p_{\pi}^+}{p_1^++p_{\pi}^+}.
\label{eqn:xi_u}
\end{equation}
    
The GPDs depend on $x_{\rm B}$, $\xi_t$ and $t$, where as the TDAs depend on $x_{\rm B}$, $xi_u$ and $u$. The $\pi^0$ production process through GPDs in the forward-angle ($t$-channel) and through TDAs in the backward-angle ($u$-channel) are schematically shown in Figs.~\ref{fig:GPD_TDA}(a) and (b), respectively. In terms of the formalism, TDAs are similar to the GPDs, except they require a switch from the impact parameter space ($t$ dependent) through Fourier transform to the large momentum transfer space ($u$ dependent).

The backward angle TDA colinear factorization scheme has similar requirements: $x_{\rm B}$ is fixed, the $u$-momentum transfer is required to be small compared to $Q^2$ and $s$; $u\equiv\Delta^2$, which implies the $Q^2$ and $s$ need to be sufficiently large. Recall the early scaling for GPD physics occurs between  $ 2 < Q^2 < 5$~GeV$^2$. The case for the backward processes was open before the pioneering studies fromJLab 6 GeV ~\cite{park18, wenliang17}. The backward $\pi^+$ and $\omega$ production results have shown indications of TDA $Q^2$-scaling at $Q^2 <<10$ GeV$^2$. Furthermore, the parameter $\Delta=p_{\pi}-p_1$ is considered to encode new valuable complementary information on the hadronic 3-dimensional structure, whose detailed physical meaning still awaits clarification~\cite{pire15}.

Beyond the JLab 12 GeV program, the backward $\pi^0$ production will be studied by the $\overline{\rm P}$ANDA experiment at FAIR~\cite{panda15}. This experimental channel can be accessed through observables including $ p + \overline{p} \rightarrow  \gamma^* + \pi^0$ and $p + \overline{p} \rightarrow J/\psi + \pi^0$. Note that this backward $\pi^0$ production involves the same TDAs as in the electroproduction case. They will serve as very strong tests of the universality of TDAs in different processes~\cite{lansberg07}.



%

\subsubsection{Further Detail on the $\pi^0 N$ TDAs}

At leading twist-3, the parameterization of the Fourier transform of the $\pi N$ transition matrix element of the three-local light cone quark operator
$\widehat{O}_{\rho \tau \chi}(\lambda_1 n, \lambda_2 n,
\lambda_3n)$~\cite{radyushkin97} can be written as~\cite{pire11}
\begin{align}
& 4 \mathcal{F} \langle \pi_{\alpha}(p_\pi)| \widehat{O}_{\rho \tau \chi}(\lambda_1 n, \lambda_2 n, \lambda_3n)| N_{\iota} (p_1) \rangle  \nonumber \\[5mm]
& = 4(P\cdot n)^3 \int \left[\, \prod^3_{j=1} \frac{d \lambda_j}{2\pi} \, \right] e^{i\sum^{3}_{k=1} x_k\lambda_k (P \cdot n)} \langle \pi_{\alpha}(p_\pi)| \widehat{O}_{\rho \tau \chi}(\lambda_1 n, \lambda_2 n, \lambda_3n)| N_{\iota} (p_1) \rangle  \nonumber \\[3mm]  
& = \delta(x_1 + x_2 + x_3 - 2\xi_u) \sum_{s.f.}\,(f_a)^{\alpha\beta\gamma}_{\iota} \, s_{\rho \tau, \chi} \, H^{\pi N}_{s.f.}(x_1, x_2, x_3, \phi, \Delta^2; \mu^2_{F})
\label{eqn:parameterization}
\end{align}
where $\mathcal{F}$ represents the Fourier transform; $P=p_1+p_\pi$ is the average $u$-channel momentum, and $\Delta=p_{\pi}-p_1$ is the $u$-channel momentum transfer, recall $\Delta^2 \equiv u$.  The spin-flavor ($s.f.$) sum over all independent flavor structure $(f_a)^{\alpha\beta\gamma}_\iota$ and Dirac structure $s_{\rho\tau,\chi}$ relevant at the leading twist; $\iota(a)$ is the nucleon (pion) isotopic index. The invariant transition amplitudes, $H^{\pi N}_{s.f.}$, which are often referred to as the leading twist $\pi N$ TDAs, are functions of the light-cone momentum fraction $x_i(i=1,2,3)$, the skewness variable $\xi_u$, the $u$-channel momentum-transfer squared $\Delta^2$, and the factorization scale $\mu_F$~\cite{lansberg11}. The full extended expression of $H^{\pi N}_{s.f.}$ can be found in Ref.~\cite{pire11}.

In a simplified notation, $H^{\pi N} (x, \xi_u, \Delta^2)$ can be written in terms of invariant amplitudes $V^{\pi N}_{1,2}$, $A^{\pi N}_{1,2}$, $T^{\pi N}_{1,2,3,4}$ ~\cite{lansberg11,pire11},
\begin{equation}
H^{\pi N}_{s.f.} = \{ V^{\pi N}_{1,2}, A^{\pi N}_{1,2}, T^{\pi N}_{1,2,3,4}\}\,.
\end{equation}
Each invariant amplitude $V^{\pi N}_{1,2}$, $A^{\pi N}_{1,2}$, $T^{\pi N}_{1,2,3,4}$ is also a function of $x_i$, $\xi_u$ and $\Delta^2$. It is important to note that not all of the $\pi N$ TDA invariant amplitude are independent~\cite{lansberg11}, and their relation are documented in Ref.~\cite{lansberg11}.

Similar to early attempts in the GPD case~\cite{ralston02}, the most straightforward solution to determine a reasonable $\Delta^2$ dependence is to perform a factorized form of $\Delta^2$ dependence for quadruple distributions. Thus, the $\pi N$ factorized form of $\Delta^2$ dependence can be written as~\cite{lansberg11}:
\begin{equation}
H^{\pi N} (x, \xi_u, \Delta^2) = H^{\pi N} (x_i, \xi_u) \times G(\Delta^2),
\label{eqn:G-Delta}
\end{equation}
where $G(\Delta^2)$ is the $\pi N$ transition form factor of the three local quarks. Note that the determination of the $\Delta^2$ dependence and extraction of the $G(\Delta^2)$ form factor will be a distant goal for backward angle physics.

\begin{figure}[t]
\centering
\includegraphics[width=0.75\textwidth]{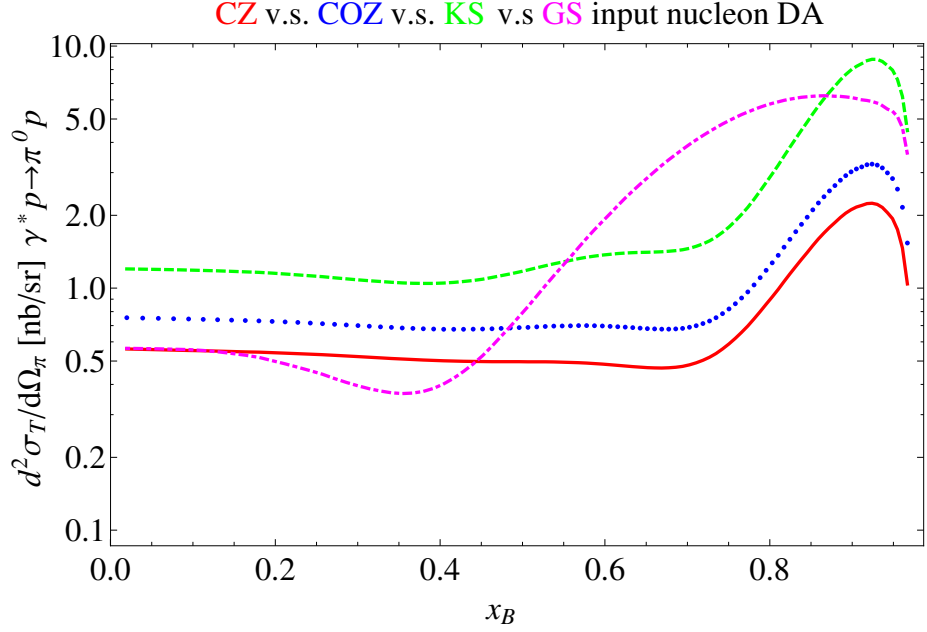} 
\caption{$d^2 \sigma_{T}/d\Omega_{\pi}$ for backward $\gamma^*p \rightarrow p\pi^0$ as a function of $x_{\textrm B}$ for $\pi N$ TDAs at $Q^2$ = 10~GeV$^2$, $u = -$0.5~GeV$^2$. CZ (solid line)~\cite{chernyak84}, COZ (dotted line) ~\cite{chernyak89}, KS (dashed line) ~\cite{king87} and GS (dash-dotted line) ~\cite{gari86} nucleon DAs were used as input. This plot was published in Ref.~\cite{lansberg11}}
\label{TDA_DA_cal}
\end{figure}

\begin{figure}[t]
\centering
\includegraphics[width=0.32\textwidth]{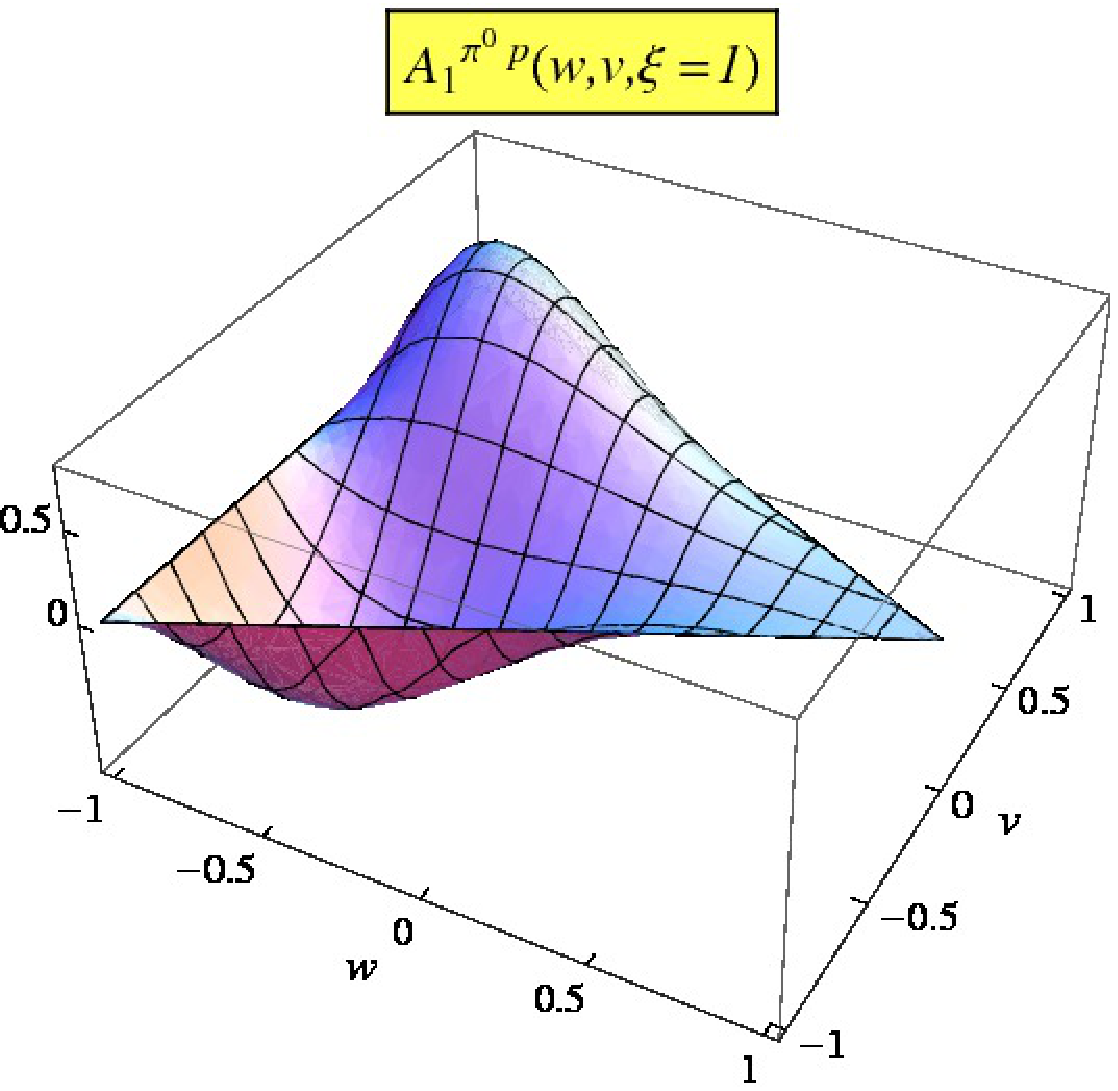}
\includegraphics[width=0.32\textwidth]{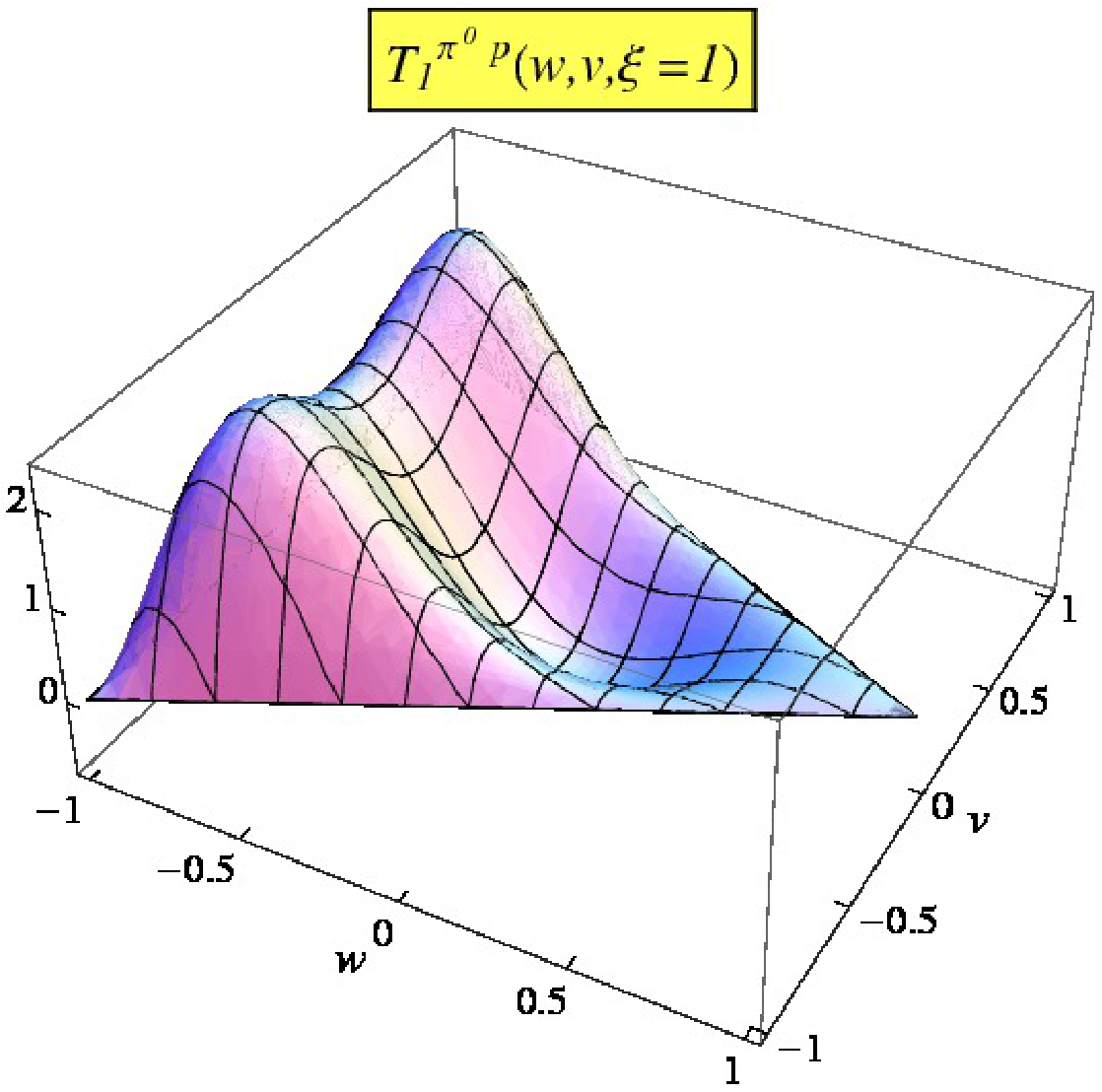}
\includegraphics[width=0.32\textwidth]{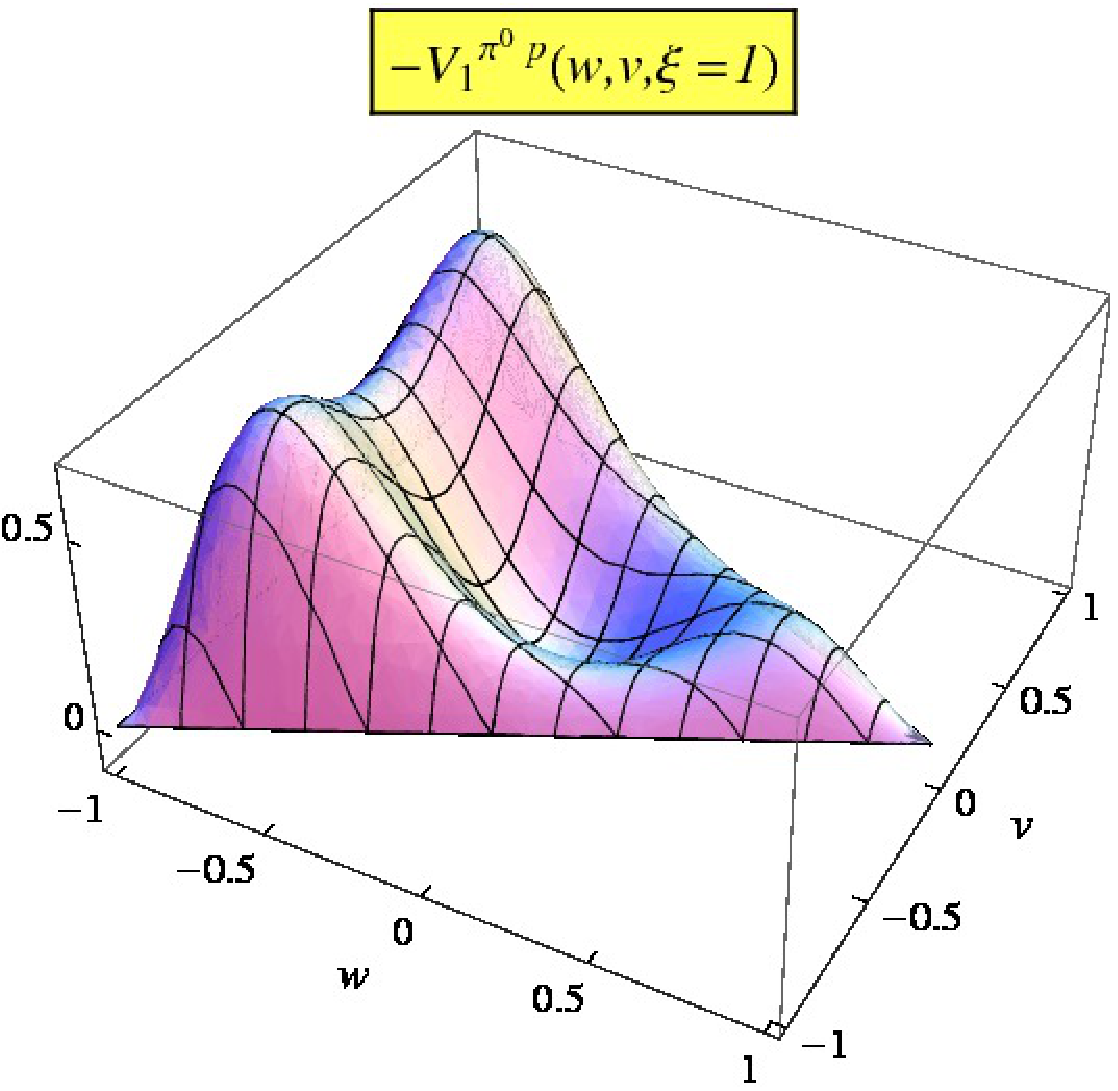}
\caption{$\pi^0p$ TDAs $V^{\pi^0 p}_1$, $A^{\pi^0 p}_1$ and $T^{\pi^0 p}_1$, computed as functions of quark-diquark coordinates, in the limit $\chi_u~\rightarrow 1$. CZ $N$ DAs are used as numerical input. These plots were published in Ref.~\cite{lansberg11}.}
\label{fig:TDA_shapes}
\end{figure}

In the $\xi_u=1$ limit, the $\pi^0 p$ TDAs: $V^{\pi^0p}_1$, $A^{\pi^0p}_1$, $T^{\pi^0p}_1$ can be simplified to the following combination of nucleon DAs~\cite{pire11}:
\begin{align}
V^{\pi^0p}_1 (x_1, x_2, x_3, \xi_u = 1) &= - \frac{1}{2} \times \frac{1}{4} \, V^P \left( \frac{x_1}{2}, \, \frac{x_2}{2}, \, \frac{x_3}{2} \right) \\[3mm]
A^{\pi^0p}_1 (x_1, x_2, x_3, \xi_u = 1) &= - \frac{1}{2} \times \frac{1}{4} \, A^P \left( \frac{x_1}{2}, \, \frac{x_2}{2}, \, \frac{x_3}{2} \right) \\[3mm]
T^{\pi^0p}_1 (x_1, x_2, x_3, \xi_u = 1) &= \frac{3}{2} \times \frac{1}{4}   \, T^P \left( \frac{x_1}{2}, \, \frac{x_2}{2}, \, \frac{x_3}{2} \right)
\end{align}
A variety of nucleon ($N$) DAs such as Chernyak-Zhitnitsky (CZ)~\cite{chernyak84}, Chernyak-Ogloblin-Zhitnitsky (COZ)~\cite{chernyak89}, King and Sachrajda (KS)~\cite{king87} and Gari and Stefanis (GS)~\cite{gari86} can be used as numerical input for $V^P$, $A^P$ and $T^P$. A TDA calculation for $\pi^0$ production cross section versus $x_{\rm B}$ is shown in Fig.~\ref{TDA_DA_cal}, where all four $N$ DAs are used.

The $N$ DA model is an important part of the TDA model prediction, and depending on the choice of the $N$ DAs the predicted experimental observables can change significantly. Therefore, improvements to the TDA parameterized formalism would rely on an accurate nucleon spectral distribution  by the $N$ DA models. In the same time, as more data are collected during JLab 12 GeV, a refined TDA model help to discriminate between different $N$ DAs. This healthy iterative process can help improving our knowledge of the proton structure~\cite{lansberg07}.

According to the TDA framework, the leading order (LO) backward angle $\gamma + p \rightarrow \pi^0 +p$ unpolarized cross section can be written as~\cite{lansberg07, lansberg11} 
\begin{equation}
\frac{d^2\sigma_{T}}{d \Omega_\pi} = |\mathcal{C}^2| \, \frac{1}{Q^6} \, \frac{\Lambda(s, m^2, M^2)}{128\,\pi^2 s (s-M^2)} \frac{1+\xi}{\xi} (|\mathcal{I}|^2 - \frac{\Delta^2_{T}}{M^2} |\mathcal{I}^\prime|^2).
\end{equation}
$\Lambda(s, m^2, M^2)$ is the Mandelstam function~\cite{lansberg11}, where $m$ corresponds to the meson mass and $M$ is the nucleon mass. In the backward angle kinematics,
\begin{equation}
\Delta^2_T = \frac{(1-\xi) \left(\Delta^2 - 2\xi \left(\frac{M^2}{1+\xi} - \frac{m^2}{1-\xi}\right)\right) }{1+\xi}\,.
\end{equation}
The coefficients $\mathcal{I}$ and $\mathcal{I}^{\prime}$ are defined as~\cite{lansberg07}
\begin{equation}
\mathcal{I} = \int\left( 2\sum^7_{\alpha=1} T_{\alpha} + \sum^{14}_{\alpha=8}T_\alpha\right), ~~~ \mathcal{I}^{\prime} = \int\left( 2\sum^7_{\alpha=1} T^{\prime}_{\alpha} + \sum^{14}_{\alpha=8}T^{\prime}_\alpha\right) 
\end{equation}
where the coefficient $T_{\alpha}$ and $T^{\prime}_\alpha (\alpha = 1, ..., 14)$ are functions of $x_i$, $y_j$, $\xi$ and $\Delta$. Here, $x_i$ and $y_j$ represent the momentum fraction for the initial and final state quark. Each of the component of $T_{\alpha}$ and $T^{\prime}$ represents one of the 21 diagrams contributing to the hard-scattering amplitude (note that the last seven diagrams are the duplications the of the first seven diagrams). 

Furthermore, $T_{\alpha}(\alpha = 1, ..., 14)$ can be written in terms of $V^{p\pi^0}_1$, $A^{p\pi^0}_1$, $T^{p\pi^0}_1$, $T^{p\pi^0}_4$ and $N$ DA ($V^p$, $A^p$, $T^p$);  $T^{\prime}_\alpha (\alpha = 1, ..., 14)$ can be written in terms of $V^{p\pi^0}_2$, $A^{p\pi^0}_2$, $T^{p\pi^0}_2$, $T^{p\pi^0}_3$ and $N$ DA~\cite{lansberg07}. This work has genuinely established the connection between the TDAs amplitudes to the cross section observables.


\subsubsection{Two Predictions from TDA Colinear Factorization}
\label{sec:tda_prediction}

The TDA colinear factorization has made two specific qualitative predictions regarding backward meson electroproduction, which can be verified experimentally~\cite{lansberg11, pire15, kirill15, pire18}:
\begin{itemize}
\item The dominance of the transverse polarization of the virtual photon results in the suppression of the $\sigma_{\rm L}$ cross section by a least ($1/Q^2$): $\sigma_{\rm L}/\sigma_{\rm T}$ $< 1/Q^2$, 
\item The characteristic $1/Q^8$-scaling behavior of the transverse cross section for fixed $x_{\rm B}$ (or at fixed $\xi_{u}$), following the quark counting rules. 
\end{itemize}
The proposed measurements in this LOI will provide significant experimental insights to challenge both of these predictions. In addition, the $-u$ dependence of the separated experimental cross section will provide insight for the extraction of the the $\pi N$ transition form factor $G(\Delta^2)$ (from Eqn.~\ref{eqn:G-Delta}).


\subsubsection{Visions of Studying Backward Physics with the TDA Framework}

As summarized in Sec.~\ref{sec:exp_summary}, JLab 6~GeV data~\cite{park16} have shown indications of $Q^2$-scaling that is consistent with the TDA prediction for $Q^2 << 10$~GeV$^2$. Therefore, it is valuable to pursue further studies using the TDA framework in the JLab 12~GeV era.

We envision a systematic study of TDAs for a given meson production process consisting of three stages:
\begin{description}

\item[Stage 1:] Study and validation of the TDA framework, by measuring the general scaling trend of the separated L/T cross sections.

\item[Stage 2:] Determination of the $\Delta^2$ dependence and the $\pi N$ transition form factor $G(\Delta^2)$ defined in Eqn.~\ref{eqn:G-Delta}.  

\item[Stage 3:] Extraction the TDAs by probing the single and double spin asymmetries for backward meson production.

\end{description}
Measurement proposed in this LOI is an important small step in the first stage of this study. 







\subsection{The Hadronic Approach}
\label{regge}

The development of Regge-trajectory-based models has created a useful linkage between physics kinematic quantities and experimental observables.  Experimental observables in the JLab physics regime are often parameterized in terms of $W$, $x_{\rm B}$, $Q^2$ and $t$. By varying a particular parameter while fixing others, one can perform high precision studies to investigate the isolated dependence of the varied parameter for a given interaction.



According the Regge models, the standard treatment to take into account the exchange of high-spin, high-mass particles is to replace the pole-like Feynman propagator of a single particle (i.e. $\frac{1}{t-M^2}$ by the Regge (trajectory) propagator. Meanwhile, the exchange process involves a series of particles of the same quantum number (following the same Regge trajectory $\alpha(t)$), instead of single particle exchange~\cite{regge60, chew62}.

\begin{figure}[ht]
\centering
\includegraphics[width=0.425\textwidth]{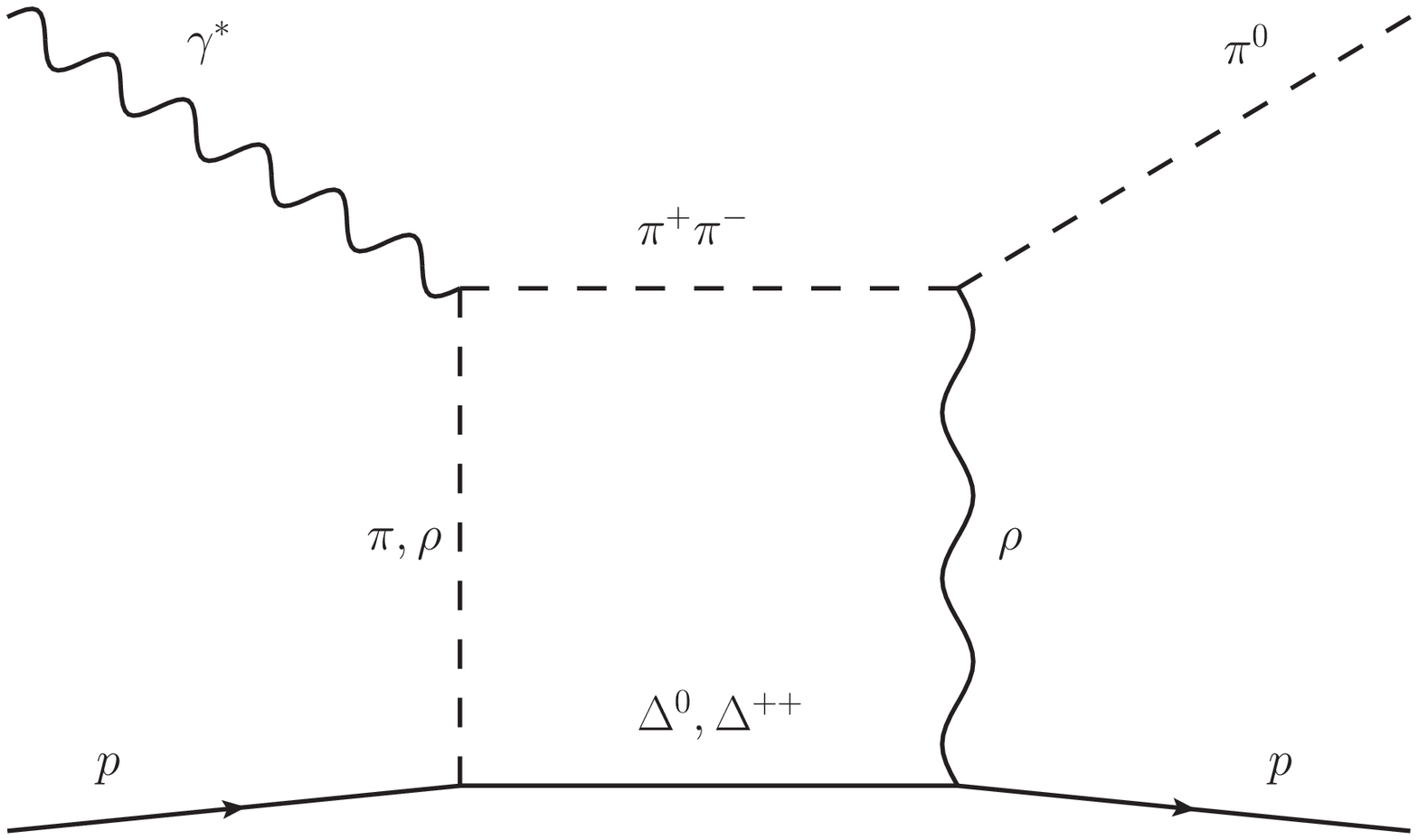}
\includegraphics[width=0.425\textwidth]{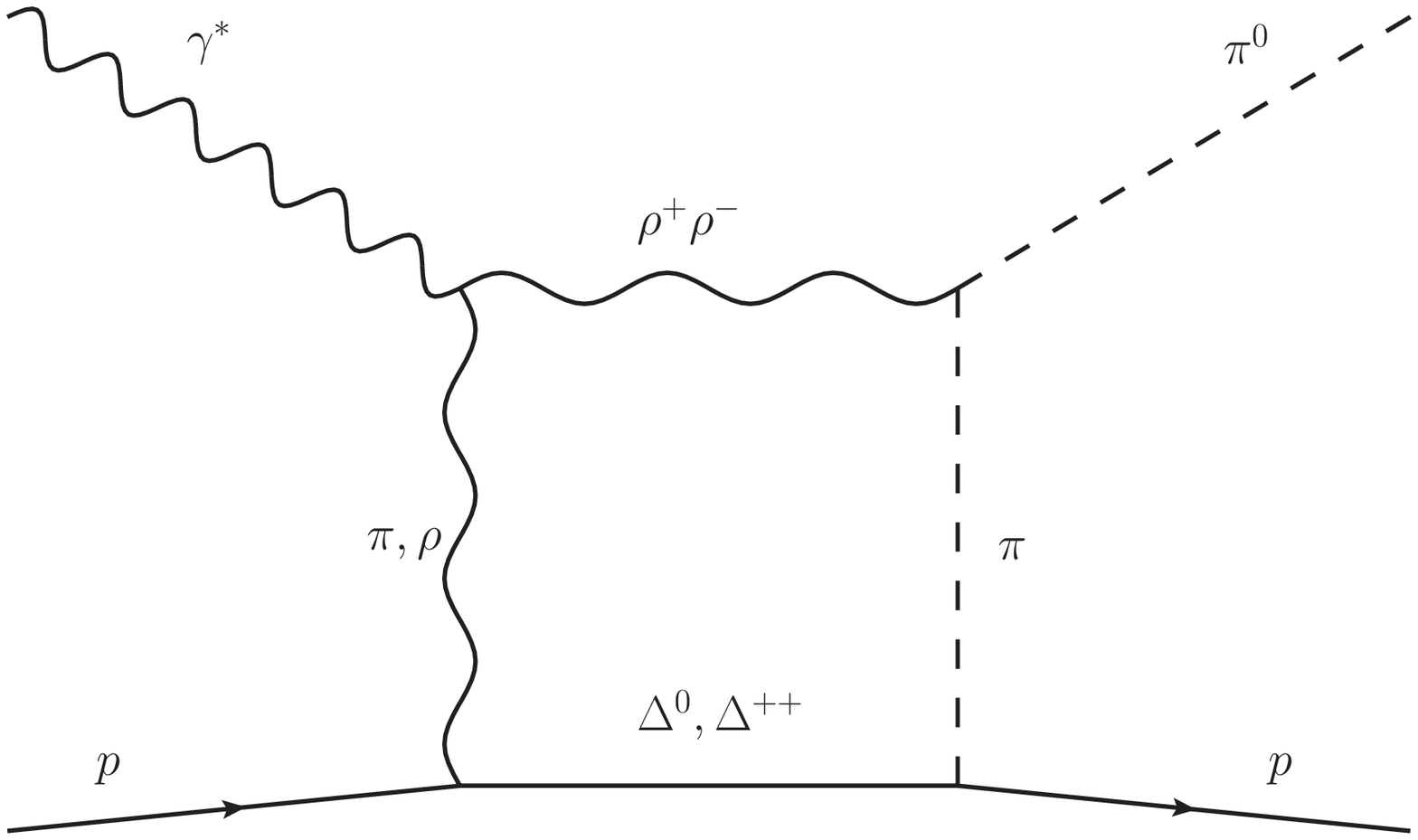}
\caption{Examples of meson exchange diagrams which contribute to forward angle $\pi^0$ production. The left plot is an example of charged $\pi$ rescattering~\cite{laget11}; the right plot is an example vector meson contribution. These plots were created based on the original ones published in Ref.~\cite{laget18}.}
\label{fig:JML_front}
~\\[4mm]
\includegraphics[width=0.425\textwidth]{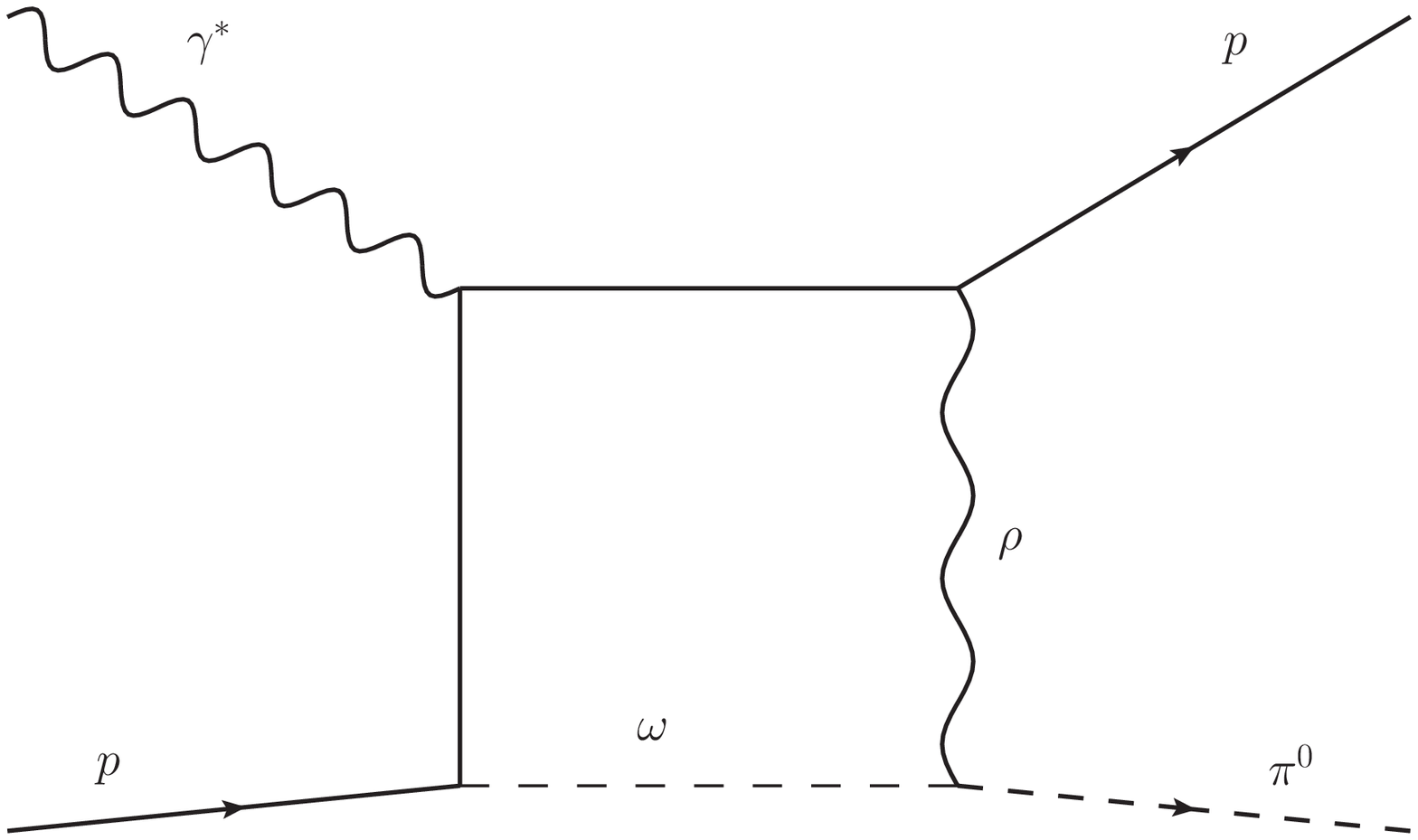}
\caption{Example of a possible meson exchange diagram which contributes to backward angle $\pi^0$ production. This plot was created based on the original one published in Ref.~\cite{laget18}.}
\label{fig:backward_pi}
\end{figure}

In the forward angle $\pi^0$ electroproduction study~\cite{laget11}, J. M. Laget linked the elastic $\pi^0$ cross section to the scattering channels of $\omega p$, $\rho^+n$, $\rho^-\Delta^{++}$, diagrams shown in Fig.~\ref{fig:JML_front}. This treatment significantly improved the prediction power of the hadronic Regge based model and led to a good agreement with the data~\cite{laget11}.

In a recent private communication~\cite{laget18}, J. M. Laget indicated that the hadronic Regge based model is capable of describing the data trend of the backward $\omega$ cross section (shown in Fig.~\ref{laget}),  at $Q^2=1.6$ and $2.45$~GeV$^2$. The preliminary conclusion from his study was that the nucleon pole contribution (baryon exchange) alone is not enough to account for the measured cross section~\cite{laget11}, thus, backward $\omega$ production required $\rho^0$, $\rho^+n$, $\rho\Delta$ scattering channels, in addition to the nucleon pole amplitude~\cite{laget18}. Note, this approach is very similar to the one used for the $\pi^0$ forward angle study. For reference purpose, a possible $u$-channel baryon trajectory exchange diagram for $\pi^0$ production is shown in Fig.~\ref{fig:backward_pi}, and this diagram is based on the knowledge of forward angle $\pi^0$ production (shown in Fig.~\ref{fig:JML_front}). Currently, a publication is in preparation which will contain more findings of the backward $\omega$ cross section using the hadronic Regge based model~\cite{laget18}.


Due to a lack of systemic studies, currently available backward angle physics data above the resonance region (most of them are summarized in Sec.~\ref{sec:exp_summary}) have limited coverage in terms of $W$, $Q^2$ and $t$ (or $u$), therefore, cannot support a full $u$-channel phenomenological study. However, it is still a useful tool to verify the key knowledge gained from the forward angle physics program, i.e. to map out the full $-t$ evolution and give the backward angle slope for a given meson production process, such as the example shown in Fig.~\ref{fig:omega}. Note that the chosen kinematic setting in the LOI is made based on the existing forward angle $\pi^0$ measurement~\cite{camsonne12}, i.e. $Q^2=3.0$ and $4.0$ GeV$^2$ at fixed $x_{\rm B}=$0.36.





\section{Experimental Methodology and Configuration for $^1$H$(e, e^{\prime}p)\pi^0$}
\label{sec:exp_kin}

\begin{figure}[t]
\centering
\includegraphics[width=0.6\textwidth]{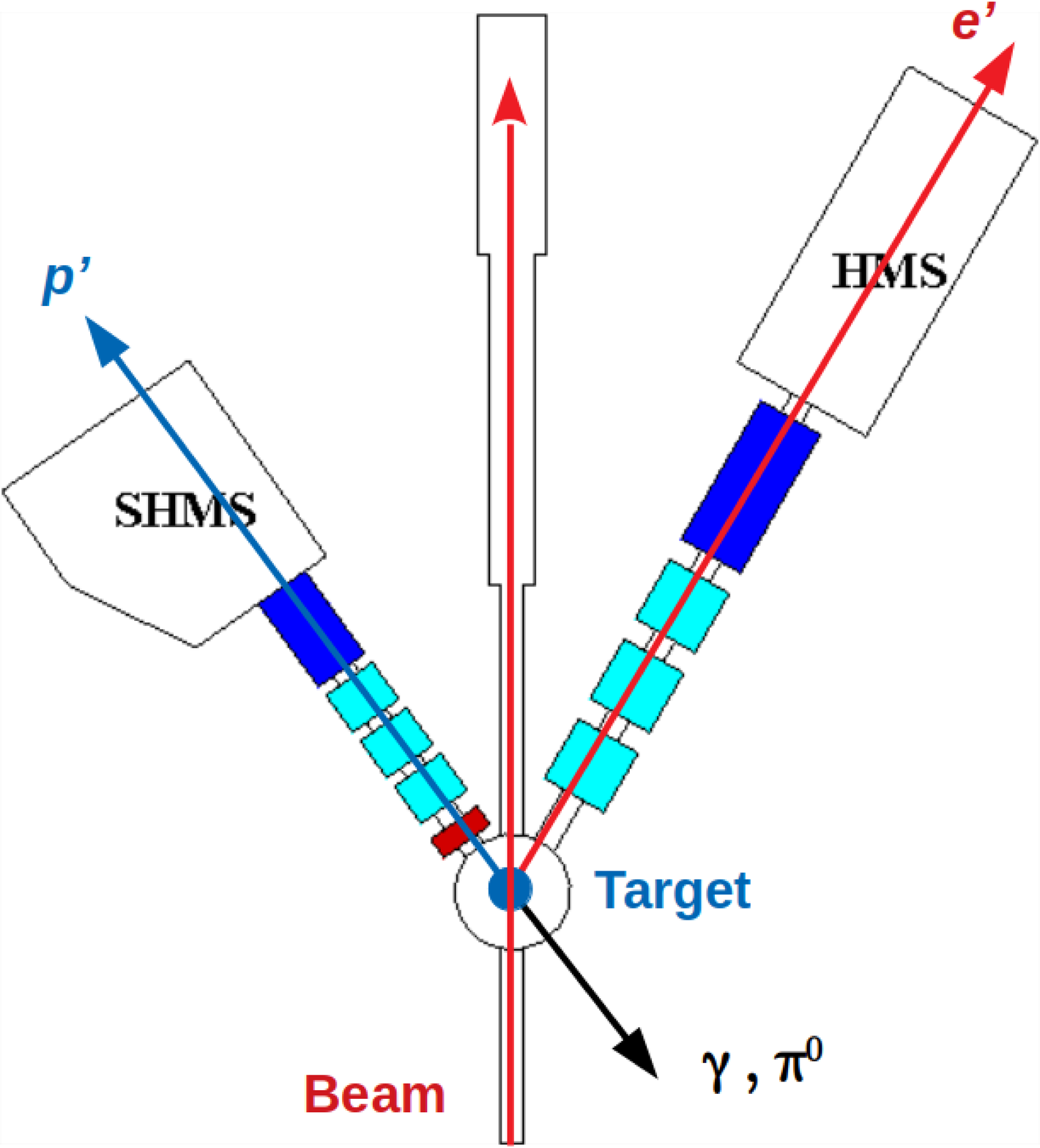}
\caption{Experimental configuration for $^1$H$(e^\prime, ep)\pi^0$ with the standard Hall C equipment. SHMS and HMS are located on the left and right side of the beam line, respectively.}
\label{fig:setup}
\end{figure}

The exclusive backward angle $\pi^0$ electroproduction measurement is proposed to use the standard Hall~C equipment: SHMS and HMS in coincidence mode, the standard-gradient unpolarized electron beam and the liquid hydrogen (LH$_2$) target. For most of settings, SHMS will be used to detect the recoiling proton and HMS will be used to detect the scattered electron. The $\pi^0$ events will be selected by using the missing mass reconstruction technique (described in Sec.~\ref{sec:missmass}. A schematic diagram of the experimental configuration for the $^1$H$(e, e^{\prime}p)\pi^0$ is shown in Fig~\ref{fig:setup}.

\subsection{The Missing Mass Reconstruction Technique}
\label{sec:missmass}

\begin{figure}[t]
\centering
\includegraphics[trim={1mm 1mm 1mm 1mm},clip,width=0.6\textwidth]{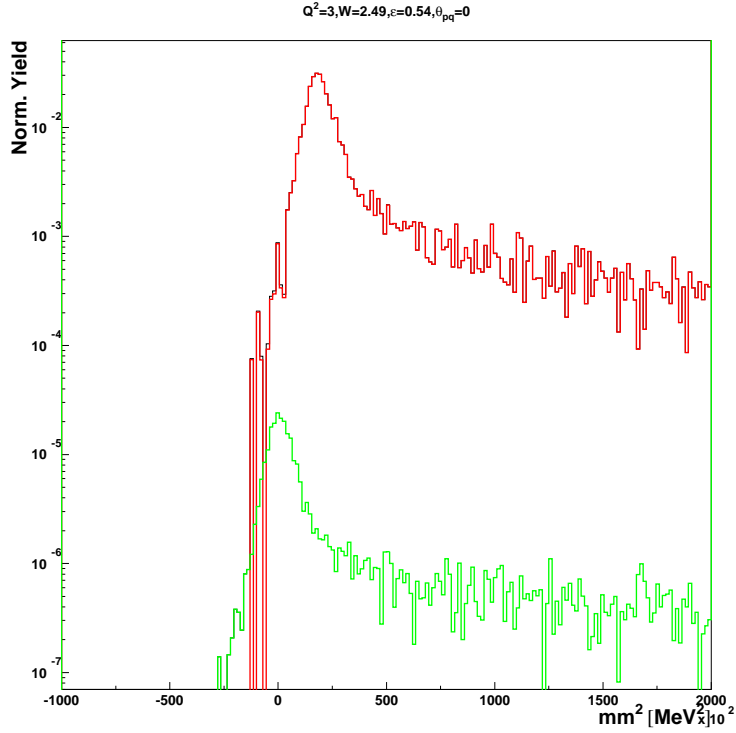}
\caption{Simulated missing mass squared distribution at $Q^2=3$~GeV. The green distribution is the backward angle DVCS contribution; the red distribution is the backward angle $\pi^0$ production. $\pi^0:\gamma$ ratio is estimated to be 1000:1. See the Appendix for details of the simulation models used.}
\label{fig:mm2}
\end{figure}

Besides its theoretical interests, the $\pi^0$ channel is chosen due to the small physics background underneath its missing mass peak. In comparison to backward angle $\omega$ electroproduction~\cite{wenliang17}, $\pi^0$ production has much less physics background from other mesons (such as $\eta$ and $\rho$). The only contributing physics background under the coincidence mode comes from the DVCS process, whose missing mass peak is near $m_x=0$~GeV. Simulated $m_x^2$ distributions for the backward $\pi^0$ and $\gamma$ are shown in Fig.\ref{fig:mm2}. The red distribution is for the $\pi^0$ events and the green distribution is for $\gamma$ events, both distributions are normalized to 1~$\mu$C of beam charge. The $\pi^0:\gamma$ production ratio is $\sim$ 1000:1 in the simple simulation models used. A $m^2_x$ cut of 90 MeV$^2$ should exclude most of the $\gamma$ events.  After events are binned in the $u$ and $\phi$, the shape and width of the $m_x$ peak will change slightly due to differences in the kinematics coverage ($Q^2$ and $W$). This will hold true for both $\pi^0$ and $\gamma$. The standard missing mass cut will not completely separate two event distributions. The Mont-Carlo simulation will be needed to estimate the $\gamma$ contamination for background subtraction purpose. This contamination should be in the order 1\% or less if one takes into account the massive difference in cross section. Note that the physics cross section model is described in Appx.~\ref{sec:pi0_MC}.

\subsection{L/T/LT/TT Separation}

The general form of two-fold differential cross section can be expressed in terms of the structure functions as: 
\begin{equation}
2 \pi \frac{d^2 \sigma}{dt ~ d\phi} = \frac{d \sigma_{\rm T}}{dt} + 
\epsilon ~ \frac{d \sigma_{\rm L}}{dt} + \sqrt{2\epsilon(1+\epsilon)}~ 
\frac{d\sigma_{\rm LT}}{dt} \cos \phi + \epsilon ~ 
\frac{d\sigma_{\rm TT}}{dt} \cos 2\phi \,.
\label{eqn:xsection_LT}
\end{equation}

\begin{figure}[t]
\centering
\includegraphics[width=0.9\textwidth]{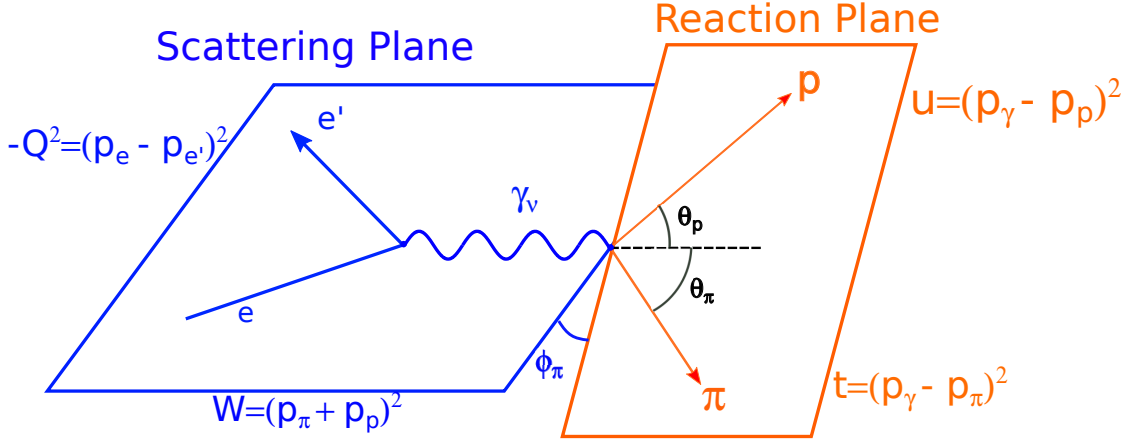}
\caption{The scattering and reaction planes for $\pi^0$ electroproduction: $^1$H$(e, e^{\prime}p)\pi^0$.  The scattering plane is shown in blue and the reaction plane is shown in orange. Note that the forward-going proton after the interaction is labelled $p$; $\gamma_\nu$ represents the exchanged virtual photon and its direction defines the $q$-vector; $\phi_p$ ($ \phi_p = \phi_{\pi} + 180^\circ$) is defined as the angle between the scattering and reaction planes (the azimuthal angle around the $q$-vector); $\theta_p$ and $\theta_{\pi}$ denote the scattering angles of the $p$ and $\pi$ with respect to the $q$-vector, respectively. The definition of the Lorentz invariant variables such as $W$, $Q^2$, $t$ and $u$ are also shown.}
\label{fig:planes}
\end{figure}

A schematic diagram of the exclusive $\pi^0$ electroproduction reaction, $^1$H$(e, e^{\prime}p)\pi^0$, giving the definition of the kinematic variables in Eqn.~\ref{eqn:xsection_LT} is shown in Fig.~\ref{fig:planes}. The three-momentum vectors of the incoming and the scattered electrons are denoted as $\vec{p}_e$ and $\vec{p}_{e^{\prime}}$, respectively. Together they define the scattering plane, which is shown as a blue box. The corresponding four momenta are p$_e$ and p$_e^\prime$. The electron scattering angle in the lab frame is labelled as $\theta_e$. The transferred four-momentum vector $q$($\nu,\vec{q}$) is defined as (p$_e -$p$_{e^\prime}$).  The three-momentum vectors of the recoil proton target ($\vec{p}_p$ ) and produced $\pi^0$ ($\vec{p}_{\pi}$) define the reaction plane, is shown as the orange box. The azimuthal angle between the scattering plane and the reaction plane is denoted by the recoil proton angle $\phi_p$. From the perspective of standing at the entrance and looking downstream of the spectrometer, $\phi_p=0$ points to horizontal left of the $q$-vector, and it follows an anticlockwise rotation. The lab frame scattering angles between $\vec{p}_p$ (or $\vec{p}_\pi$) and $\vec{q}$ are labeled $\theta_p$ (or $\theta_\pi$). Unless otherwise specified, the symbols $\theta$ and $\phi$ without subscript are equivalent to $\theta_p$ and $\phi_p$, since the recoil protons will be detected during the experiment. The parallel and antiparallel kinematics are unique circumstances, and occur at $\theta = 0^\circ$ and $\theta = 180^\circ$, respectively. 


The Rosenbluth separation, also known as the longitudinal/transverse (L/T) separation, is a unique method of isolating the longitudinal component of the differential cross section from the transverse component. The method requires at least two separate measurements with different experimental configurations, such as the spectrometer angles and electron beam energy, while fixing the Lorentz invariant kinematic parameters such as $x_{\rm B}$ and $Q^2$. The only physical parameter that is different between the two measurements is $\epsilon=\left(1+2\frac{|\vec{q}|^2}{Q^2}\tan^2\frac{\theta}{2}\right)^{-1}$, which is directly dependent upon the incoming electron beam energy ($E_{e^\prime}$) and the scattering angle of the outgoing electron.

\begin{figure}[t]
\centering
\includegraphics[width=0.8\textwidth]{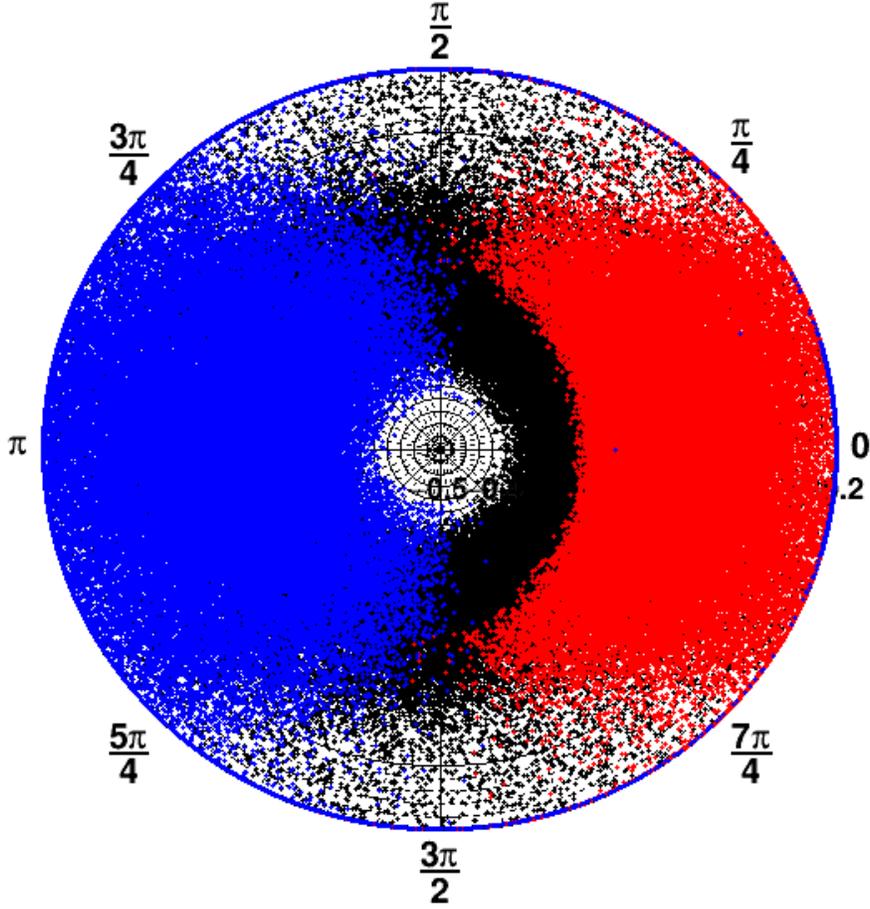}
\caption{$u'$-$\phi$ polar distributions at $Q^2 = 2$ GeV$^2$ and $\epsilon=0.52$. $-u$ is plotted as the radial variable and $\phi$ as the angular variable. The blue points represent data at $\theta_{pq}=+3^\circ$, black points represent data at $\theta_{pq} = 0^\circ$, and red data points represent data at $\theta_{pq} = -3^\circ$. The center of the plot represent $-u=-0.5$~GeV$^2$ and the outer circle is at $-u=0.2$~GeV$^2$.}
\label{fig:bullseye}
\end{figure}

Even though the SHMS setting at $\theta_{pq}=0$ is centered with respect to the $q$-vector, corresponding to the parallel scenario for the proton (anti-parallel for $\pi$), the spectrometer acceptance of the SHMS (proton arm) is not wide enough to provide uniform coverage in $\phi$ (black events in Fig.~\ref{fig:bullseye}). A complete $\phi$ coverage over a full $u$ range is critical for the extraction of the interference terms (LT and TT) during the L/T separation procedure. To ensure an optimal $\phi$ coverage, additional measurements are required at the $\theta=\pm3~\circ$ SHMS angles (blue and red events). Constrained by the minimum SHMS angle from the beam line of $\theta_{\rm SHMS} = 5.5^\circ$, the lower $\epsilon$ measurement is only possible at two angles at each $Q^2$. However, this can be compensated by the full $\phi$ coverage at the higher $\epsilon$ measurement and the simulated distribution, thus determining the interference components (LT and TT) of the differential cross section.

The last step of the L/T separation is to fit the experimental cross section versus $\phi$ for a given $u$ bin. The lower and higher epsilon data will be fitted simultaneously using Eq.~\ref{eqn:xsection_LT} to ensure successful extraction of the $\sigma_{\rm T, L, LT, TT}$. The common offset between and difference between the lower and higher $\epsilon$ data set give raise to the $\sigma_{\rm T}$ and $\sigma_{\rm L}$; whereas the $\phi$ dependence signifies the $\sigma_{\rm LT}$ and $\sigma_{\rm TT}$ contribution.

\subsection{Proposed Kinematics}
\label{sec:kin}

\begin{table}[t]
\centering
\setlength{\tabcolsep}{0.38em}
\caption{Proposed kinematics for the $^1$H$(e, e^\prime p)\pi^0$ measurement. Note that the $W$ and $Q^2$ are the same for the $^1$H$(e, e^\prime p\gamma)$ reaction. For most of the settings, HMS will detect the scattered electron and the SHMS will detect the recoiled proton. For $Q^2=2$ GeV (indicated by $^*$), the SHMS will detect the electron and the HMS will detect the proton because the scattered electron momentum and angle at high $\epsilon$ are too high and too far forward for the HMS. Note that at $Q^2 = 3$ and $4$ GeV$^2$, E12-13-010 will provide the L/T separated cross section at $t^{\prime} \sim 0$~\cite{E12-13-010}. }
\label{tab:kinematics}
\begin{tabular}{cccccccccccc}
\toprule
$Q^2$   & $W$  & $x_{\rm B}$  &  $E_{\rm Beam}$ & $\epsilon$ & $\theta_{\rm HMS}$  & $P_{\rm HMS}$  & $\theta_{\rm SHMS}$  & $P_{\rm SHMS}$ & $\theta_{pq}$ & $u^{\prime}$  & $-t$     \\
GeV$^2$ & GeV  &      &  GeV        &            & Degree                        & GeV/c  & Degree          & GeV/c  & Degree & GeV$^2$       & GeV$^2$  \\ 
\toprule
2.0   & 2.11 & 0.36 & 4.4$^*$    & 0.52$^*$  & 13.71$^*$  & 3.51$^*$  & $-$32.60$^*$   & $-$1.44$^*$   & 0.0$^*$  &  0.0$^*$ & 5.05$^*$  \\
      &      &      & 10.9$^*$   & 0.94$^*$ & 21.54$^*$ &  3.51$^*$    & $-$8.72$^*$   & $-$7.94 $^*$  & 0.0$^*$  &  0.0$^*$ & 5.05$^*$   \\
\hline
3.0   & 2.49  & 0.36 &  6.60        & 0.54       & 26.50    & $-$2.17   & $-$11.70      &  5.00    & 0.0  &  0.0 & 7.79    \\
       &       &      & 10.90        & 0.86       & 11.80    & $-$4.37   & $-$16.20      &  5.00    & 0.0  &  0.0 & 7.79    \\
\hline
4.0   & 2.83  & 0.36 &  8.80        & 0.55       & 22.89    &  $-$2.89   & $-$10.35      &  6.50   & 0.0   &  0.0 & 10.56    \\
       &       &      &  10.90       & 0.73       & 15.59    &  $-$4.99   & $-$12.39      &  6.50   & 0.0   &  0.0 & 10.56    \\
\hline
5.5     & 3.26  & 0.36  &  10.90        & 0.45       & 24.62    &  $-$2.78   & $-$7.86   &  8.72   & 0.0 &  0.0 & 14.69    \\
\hline
\toprule
\end{tabular}

\end{table}

\begin{figure}[h]
\centering

\includegraphics[width=0.76\textwidth]{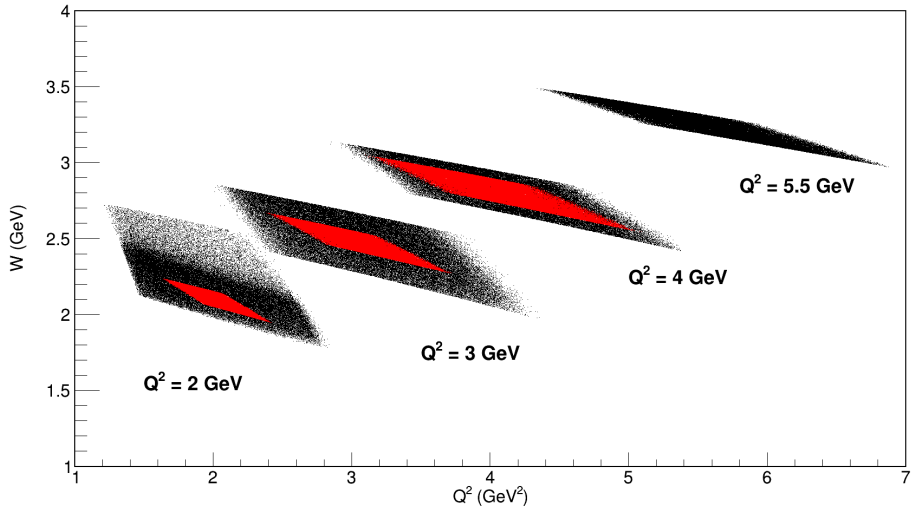}
\caption{$W$ vs $Q^2$ diamonds for the $Q^2=2.0$, $3.0$, $4.0$ and $5.5$ GeV$^2$ settings. The black diamonds are for the higher $\epsilon$ settings and the red diamonds are for the lower $\epsilon$ settings. The overlap between the black and red diamond is critical for the L/T separation at each setting. The boundary of the low $\epsilon$ (red) data coverage will become a cut for the high $\epsilon$ data. Note that the $Q^2=5.5$ GeV$^2$ setting has measurement at only one $\epsilon$ value. }
\label{fig:diamond}
\end{figure}

The $^1$H$(e, e^{\prime}p)\pi^0$ experimental yield will be measured at $Q^2=2.0$, $3.0$, $4.0$ and $5.5$ GeV$^2$, as listed in Table~\ref{tab:kinematics}.  We intend to perform L/T/LT/TT separations for all except the $Q^2=5.5$ GeV$^2$ setting.


$W$ versus $Q^2$ distributions (the `diamond' distributions) for all settings are shown in Fig.~\ref{fig:diamond}. The L/T separated cross sections at $Q^2=2.0$, $3.0$ and $4.0$~GeV$^2$ will provide the $-u$ dependence for $\sigma_{\rm L}$ and $\sigma_{\rm T}$, in addition to the behavior of $\sigma_{\rm L}$/$\sigma_{\rm T}$  ratio as function of $Q^2$.  The $Q^2=5.5$~GeV$^2$ setting is specifically to test the $Q^2$ scaling nature of the unseparated cross section.


\subsection{Projected Rates}

\begin{table}[t]
\centering
\caption{Estimated event rate (per hour) and the time duration for collecting 1000 events. These estimations take into account cuts such as the spectrometer acceptance cut, missing mass cut and the diamond cut (see Fig.~\ref{fig:diamond}). The estimated total beam time for the $\pi^0$ measurement is 1200 (PAC) hours.}
\label{tab:rates}

\setlength{\tabcolsep}{0.4em}

%

\begin{tabular}{cccccc}
\toprule 
Q$^{2}$ & W & $\epsilon$ & Rate & Time (hour) for    &  Time     \\
 & & & (per hour)        & 1000 events               &  (hours)  \\
\hline                                                                              
2.0 & 2.11 & 0.52          & 55.8  & 18              &  144    \\
    &      & 0.94          & 523.8 & 2               &  16     \\
\hline                                                                              
3.0 & 2.49 & 0.54          & 29.0  & 35              &  280    \\
    &      & 0.86          & 147.6 & 7               &   56    \\
\hline                                                                              
4.0 & 2.83 & 0.56          & 34.8  & 29              &  232    \\
    &      & 0.73          & 66.8  & 15              &  120    \\
\hline                                                                              
5.5 & 3.26 & 0.44          &  6    &  166            &  332    \\
\toprule

\end{tabular}

\end{table}


Using the backward $^1$H$(e, e^\prime p)\pi^0$ and $^1$H$(e, e^\prime p)\gamma$ physics models (see Appx.~\ref{sec:pi0_MC} for a detailed description), the estimated event rates and times for collecting 1000 events are presented in Table~\ref{tab:rates}.  In order to ensure the maximum $\phi$ coverage, each $Q^2$-$\epsilon$ point requires three proton spectrometer (SHMS in most cases) angle settings: left ($\theta_{pq}=-3^\circ$), center ($\theta_{pq}=0^\circ$) and right ($\theta_{pq}=+3^\circ$) with respect to the $q$-vector, as shown in Fig.~\ref{fig:bullseye}. The total time required for each $Q^2$-$\epsilon$ setting are listed in Table~\ref{tab:rates},  this time will be shared equally by the three angle settings. The time required to complete the $\pi^0$ measurement is 1200 (PAC) hours.

The estimated time for the $Q^2=5.5$~GeV setting is 166 hours per 1000 events, therefore we will take a close look at this particular setting when preparing the final proposal. In addition, parasitic data available from the upcoming kaon form factor experiment (E12-09-011 \cite{12-09-011}) at $Q^2=5.5$~GeV$^2$ will provide additional insight to the experimental rate expectation for this setting.


\section{Possibility of Accessing the Backward Angle DVCS}
\label{sec:DVCS}

DVCS is an important reaction, since it offers the cleanest access to multiple GPDs. One of the significant advantages of studying backward angle DVCS compared to its forward angle counterpart is the suppression of the Bethe-Heitler contribution~\cite{lansberg07}. In addition, one can test the universality of the TDA framework through comparisons of the experimental cross sections in $p+\overline{p}\rightarrow \gamma + \gamma$ at $\bar{\rm P}$ANDA and the electroproduction at JLab 12~GeV~\cite{lansberg07}.

The PAC approved DVCS experiments ~\cite{camsonne12, camsonne06} are an important part of the 12 GeV program. However, the kinematic coverage of these measurements do not include the backward angle region ($u \approx u_{min}$). Hall C, with its unique figures in terms of resolution, luminosity and detector configuration, has the potential to perform high precision measurement in this unexplored kinematic region.

As shown in Fig.~\ref{fig:mm2}, the SHMS and HMS in coincidence mode is capable of resolving the  $\pi^0$ peak, however, the DVCS peak is overwhelmed by the tail of $\pi^0$ at $m^2_X = 0$~GeV$^2$. Therefore the backward scattered real photon must be detected to study the backward DVCS. Experimentally, these backward photons can be detected by using the Neutral Particle Spectrometer (NPS)~\cite{tanja15} that is currently under construction. In addition, the recoiled proton and electrons will be detected by SHMS-HMS in coincidence mode to ensure exclusivity: $^1$H$(e, e^{\prime}p\gamma)$. The $W$ and $Q^2$ coverage for the DVCS is the same as the one for the $\pi^0$, shown in Table~\ref{tab:kinematics}.

The experimental configuration for detecting the DVCS events are listed as the following: 
\begin{itemize}
\item Triple coincidence configuration: SHMS-HMS-NPS. Where the NPS is sitting $180^\circ$ backwards of the proton spectrometer. 
\item Backward scattered photon need to be detected in the energy range between 200-500~MeV.
\item Missing mass ($M_X$) resolution of 8~MeV/$c^{2}$.
\end{itemize}

In addition, we foresee the following experimental challenges:
\begin{itemize}
\item A new movable 5-ton stand is required to host the NPS in the backward angle region.
\item The NPS designed specification has a detection efficiency of 95\% and energy resolution of 8~MeV for photon energies greater than 500~MeV. Therefore, the efficiency and resolution of low energy photon detection with the NPS, requires further study.
\end{itemize}

In short, the backward DVCS is a non-trivial and challenging measurement, which demands significant resources to accomplish. However, the scientific outcome is extremely rewarding. It is our hope to receive the PAC's feedback and comments on the DVCS part of the proposal separately from the $\pi^0$ part. A full experimental proposal will be developed accordingly based PAC's comments.

%
%
%
%

\section{Closing Remarks}

In this LOI, we present a vision for initiating a backward angle electroproduction program in the JLab 12 GeV era.

The $\pi_0$ component of the proposal will explore the transition from meson-nucleon to quark-hadron degrees of freedom using the hadronic Regge based and QCD GPD-like TDA approaches.  The estimated beamtime required for this study is 1200 PAC hours.  Careful estimations of the final uncertainties on the L/T separated cross sections and use the parasitic data from already approved proposals to refine the rate estimations are still required.  The systematic checks such as luminosity scans and $^1$H$(e, e^{\prime})p$ runs will be designed and included in the final proposal.

As stated in Sec.~\ref{sec:DVCS}, the DVCS component of the experiment requires additional detector components, resources and a more detailed feasibility study. Depending as PAC's feedback, the DVCS component will either be included or separated from the $\pi_0$ component of the proposal.


\appendix

\section{Monte Carlo model of Deep Exclusive $\pi^0$ Production}

\label{sec:pi0_MC}

The Monte Carlo studies needed for this proposal require a reaction model for an experimentally unexplored region of kinematics.  This appendix describes the model and the constraints used. The differential cross section for exclusive $\pi$ production from the nucleon can be written as
\begin{equation}
  \frac{d^{5} \sigma}{dE' d\Omega_{e'} d\Omega_{\pi}} = \Gamma_{V} \frac{d{^2}
  \sigma}{d\Omega_{\pi}}.
\end{equation}
The virtual photon flux factor $\Gamma_{V}$ is defined as
\begin{equation}
  \Gamma_v=\frac{\alpha}{2\pi^2} \frac{E'}{E} \frac{K}{Q^2}\frac{1}{1-\epsilon},
\end{equation}
where $\alpha$ is the fine structure constant, $K$ is the energy of real photon equal to the photon energy required to create a system with invariant mass equal to $W$ and $\epsilon$ is the polarization of the virtual photon.
\begin{equation}
  K=(W^2-M_p^2)/(2 M_p)
\end{equation}
\begin{equation}
  \epsilon=\left(1+\frac{2 |\mathbf{q}|^2}{Q^2} \tan^2\frac{\theta_{e}}{2}
  \right)^{-1},
\end{equation}
where $\theta_{e}$ is the scattering angle of scattered electron.

The two-fold differential cross section $\frac{d{^2} \sigma}{d\Omega_{\pi}}$ in the lab frame can be expressed in terms of the invariant cross section in center of mass frame of the photon and nucleon,
\begin{equation}
  \frac{d^2 \sigma}{d\Omega_\pi}= J \frac{d^2 \sigma}{dt d\phi},
\end{equation}
where $J$ is the Jacobian of transformation of coordinates from lab $\Omega_{\pi}$ to $t$ and $\phi$ (CM). 

In the one-photon exchange approximation, the unpolarized nucleon cross section for $n(e,e^{\prime}\pi^{-})p$
can be expressed in four terms. Two terms correspond to the polarization states of the virtual photon (L and T) and two states correspond to the interference of polarization states (LT and TT),
\begin{equation}
  d\sigma_{UU} =  \epsilon  \frac{d\sigma_{\mathrm{L}}}{dt}
  + \frac{d\sigma_{\mathrm{T}}}{dt} + 
  \sqrt{2\epsilon (\epsilon +1)} \frac{d\sigma_{\mathrm{LT}}}{dt} \cos{\phi}
  + \epsilon  \frac{d\sigma_{\mathrm{TT}}}{dt} \cos{2 \phi},
  \label{eqn:cross-2}
\end{equation}
where $\phi$ is the angle between lepton plane and hadron plane (Fig.~\ref{fig:planes}). The first two terms of Eqn.~\ref{eqn:cross-2} correspond to the polarization states of the virtual photon (L and T) and last two terms correspond to the interference of polarization states (LT and TT). $\epsilon$ is the ratio of longitudinal to transverse virtual-photon fluxes
\begin{equation}
  \epsilon=\left(1+\frac{2
  |\mathbf{q}|^2}{Q^2} \tan^2\frac{\theta_{e}}{2} \right)^{-1}.
\end{equation}

The following data and calculations were used as constraints on the parameterizations used in this model:
\begin{itemize}
\item
From Hall A, $L/T/LT/TT$ separated experimental data of exclusive electroproduction of $\pi^0$ on $^1$H are available at $x_{\rm B}=$0.36 and three different $Q^2$ values ranging from 1.5 to 2 GeV$^2$ \cite{defurne16}. Of these three, we use only the data set at $Q^2$=1.75 GeV$^2$, as it spans the widest $t$-range, $0.184<-t<0.284$ GeV$^2$~\cite{defurne16}.
\item
A GPD-based handbag-approach calculation by Goloskokov and Kroll \cite{gk11} for the E12-13-010 proposal \cite{E12-13-010} at $x_{\rm B}=$0.36, $Q^2$=3.0, 4.0, 5.5 GeV$^2$~\cite{gk11}.

\end{itemize}

Since both of these data and calculations are for forward-angle kinematics, we used the following prescription to obtain a crude model for the unique backward-angle kinematics proposed here.  
\begin{itemize}
\item
The $t$-dependence of the T/LT/TT structure functions at each $Q^2$ were fitted with functions of the form $a+b/(-t)$, which gave good fits over the range $-t_{min}<-t<0.8$ GeV$^2$ with a minimum of fit parameters. $\sigma_L$ displayed very little $t$-dependence over the region for which there was data, so it was simply taken as a small constant value with $t$ (about 1~nb/GeV$^2$, but with magnitude dropping as $Q^2$ increases).
\item
Since the electroproduction data in Fig.~\ref{fig:omega} display a forward to backward angle peak ratio of about 10:1, we estimate the magnitude of the backward angle cross sections by switching the $u$-slope for $t$-slope in the above equations, and divide by ten.
\item
Linear interpolation was performed between the parameterized values at fixed $Q^2$=1.75, 3.0, 4.0, 5.5 GeV$^2$ to obtain the L/T/LT/TT cross sections for the exact $Q^2$ needed for each event in the SIMC Hall C Monte Carlo
simulation.
\item
After the parametrization of $\sigma_{L,T,LT,TT}$ for $-u$ and $Q^2$, we assume the same $W$ dependence as used in \cite{blok08} for exclusive $\pi^+$ electroproduction at similar $x_{\rm B}$, which is $(W^2-M^2)^{-2}$ where $M$ is the proton mass.
\end{itemize}

Clearly, this model can only be described as a `best guess' of the actual DEMP $\pi^0$ cross sections in this unexplored regime.  It is anticipated that some parasitic $\pi^0$ backward angle data will be acquired in the upcoming Hall C DEMP experiments, E12-09-011, E12-06-101, E12-07-105, which can be used to improve the crude model used here.

\subsection{Exclusive $\gamma$ model}

The backward-angle DVCS model is similar in nature.  It is based on the Kumericki and Mueller KM15 model at $x_{\rm B}$=0.36, which includes the 2015 CLAS and Hall A DVCS data in fixing the GPD parameters, and uses the same prescription to convert from forward-angle to backward-angle kinematics.  No correction for Bethe-Heitler background is made, as the Bethe-Heitler process is strongly suppressed in backward kinematics and may be safely neglected \cite{pire05}. We are working on an improved DVCS model, but it is not ready in time for this LOI.

\end{document}